\documentclass[apj]{emulateapj}

\usepackage{float}
\usepackage{epsfig}
\usepackage{comment}
\usepackage{amsmath}
\usepackage{mathtools}

\long\def\symbolfootnote[#1]#2{\begingroup%
\def\thefootnote{\fnsymbol{footnote}}\footnote[#1]{#2}\endgroup}

\def\figtxt{\footnotesize}
\def\myfont{\fontsize{10}{10} \selectfont}

\def\Om{$\Omega_M$}
\def\Olam{$\Omega_{\Lambda}$}
\def\um{$\mu$m}

\def\x{$\times$}
\def\twid{$\sim$}

\def\figtxt{\footnotesize}

\def\z{redshift}
\def\vmax{$V_{{\rm max}}$}
\def\zf{ZFOURGE}

\def\mstar{$M^*$}
\def\phistar{$\Phi^*$}
\def\msol{M$_{\odot}$}
\def\msun{M$_{\odot}$}
\def\logm{Log(M/\msol )}
\def\Qui{Quiescent}
\def\qui{quiescent}
\def\rhotot{$\rho$}

\def\sch{Schechter}

\input Starburst.fd

\begin{document}

\title{
Galaxy Stellar Mass Functions from ZFOURGE/CANDELS: \\
An Excess of Low-Mass Galaxies Since \MakeLowercase{z}=2 and the Rapid
Buildup of Quiescent Galaxies\footnotemark[*] 
}

\author{Adam R. Tomczak\footnotemark[1]\footnotemark[$\dagger$],
Ryan F. Quadri$^{2,3}$
Kim-Vy H. Tran$^{1}$, 
Ivo Labb\'e\footnotemark[4],
Caroline M. S. Straatman$^4$,
Casey Papovich$^1$,
Karl Glazebrook$^5$,
Rebecca Allen$^{5,6}$,
Gabriel B. Brammer$^7$,
Glenn G. Kacprzak$^{5,8}$,
Lalitwadee Kawinwanichakij$^1$,
Daniel D. Kelson$^2$,
Patrick J. McCarthy$^2$,
Nicola Mehrtens$^1$,
Andrew J. Monson$^2$,
S. Eric Persson$^2$,
Lee R. Spitler\footnotemark[5],
Vithal Tilvi$^1$,
Pieter van Dokkum$^9$
}
\footnotetext[*]{ This paper includes data gathered with the 6.5 meter Magellan
  Telescopes located at Las Campanas Observatory, Chile. }
\footnotetext[1]{ George P. and Cynthia W. Mitchell Institute for
Fundamental Physics and Astronomy, Department of Physics and
Astronomy, Texas A\&M University, College Station, TX 77843 }
\footnotetext[2]{ Carnegie Observatories, Pasadena, CA 91101, USA } 
\footnotetext[3]{ Hubble Fellow }
\footnotetext[4]{ Sterrewacht Leiden, Leiden University, NL-2300 RA
  Leiden, The Netherlands }
\footnotetext[5]{ Centre for Astrophysics \& Supercomputing, Swinburne
University, Hawthorn, VIC 3122, Australia }
\footnotetext[6]{Australian Astronomical Observatories, PO Box 915
North Ryde NSW 1670, Australia}
\footnotetext[7]{European Southern Observatory, Alonso de C\'ordova
  3107, Casilla 19001, Vitacura, Santiago, Chile}
\footnotetext[8]{ Australian Research Council Super Science Fellow}
\footnotetext[9]{ Department of Astronomy, Yale University, New Haven,
CT 06520, USA }
\footnotetext[$\dagger$]{tomczak@physics.tamu.edu}

\begin{abstract} \myfont

Using observations from the FourStar Galaxy Evolution Survey
(\zf), we obtain the deepest measurements to date of the galaxy
stellar mass function at $0.2 < z < 3$.
\zf\ provides well-constrained photometric \z s made possible through deep medium-bandwidth imaging at
1--2\um .
We combine this with \emph{HST} imaging from the Cosmic Assembly
Near-IR Deep Extragalactic Legacy Survey (CANDELS), allowing for
the efficient selection of both blue and red galaxies down to stellar
masses $\sim 10^{9.5}$\msun\ at $z \sim 2.5$.
The total surveyed area is 316 arcmin$^2$ distributed over three
independent fields.
We supplement these data with the wider and shallower NEWFIRM
Medium-Band Survey (NMBS) to provide stronger constraints at
high masses.
Several studies at $z \leq 1.5$ have revealed a steepening of the slope
at the low-mass end of the stellar mass function (SMF), leading to an
upturn at masses $< 10^{10}$\msun\ that is not well-described by a standard single-\sch\ function.
We find evidence that this feature extends to at least $z \sim 2$, and
that it can be found in both the star-forming and \qui\ populations individually.
The characteristic mass (\mstar) and slope at the lowest masses ($\alpha$)
of a double-\sch\ function fit to the SMF stay roughly constant at \logm\
$\sim 10.65$ and $\sim -1.5$ respectively. 
The SMF of star-forming galaxies has evolved primarily in
normalization, while the change in shape is relatively minor.
Our data allow us for the first time to observe a rapid buildup at the
low-mass end of the \qui\ SMF. 
Since $z=2.5$, the total stellar mass density of \qui\ galaxies (down
to $10^9$\msun ) has increased by a factor of \twid 12
whereas the mass density of star-forming galaxies only increases by a
factor of \twid 2.2.

\end{abstract}
\maketitle

\section{Introduction}

Galaxy formation and evolution depend on the physics
governing dark matter, baryons, and interactions between the two. 
The process starts with the collapse of dark
matter halos out of the initial density perturbations in
the early universe. 
As halos continue to merge and grow they accrete gas, converting it to stars forming the
stellar mass of a galaxy. A variety of feedback processes are known to inhibit star formation, but these processes are poorly understood and can generally only be observed indirectly.


These effects in combination dictate the growth of a galaxy's stellar
mass. One of the most fundamental ways to trace these effects is to
measure the evolution of the galaxy stellar mass function (SMF) over
cosmic time.  It is well-known that the SMF does not follow the mass
function of dark matter halos; this disagreement points to differences
in the pathways that galaxies accumulate stellar mass and dark matter.
Thus, measurements of the SMF provide constraints on the feedback
processes that regulate star formation. Much work has gone into
measuring the SMF in recent years, and the development of deep near-IR
surveys has allowed these studies to push to higher redshifts and
to lower stellar masses
\citep[e.g.][]{Perez08,Drory09,Marchesini09,Ilbert10,Brammer11,Santini12,Moustakas13,Muzzin13}.

\begin{figure*}[t]
\epsfig{ file=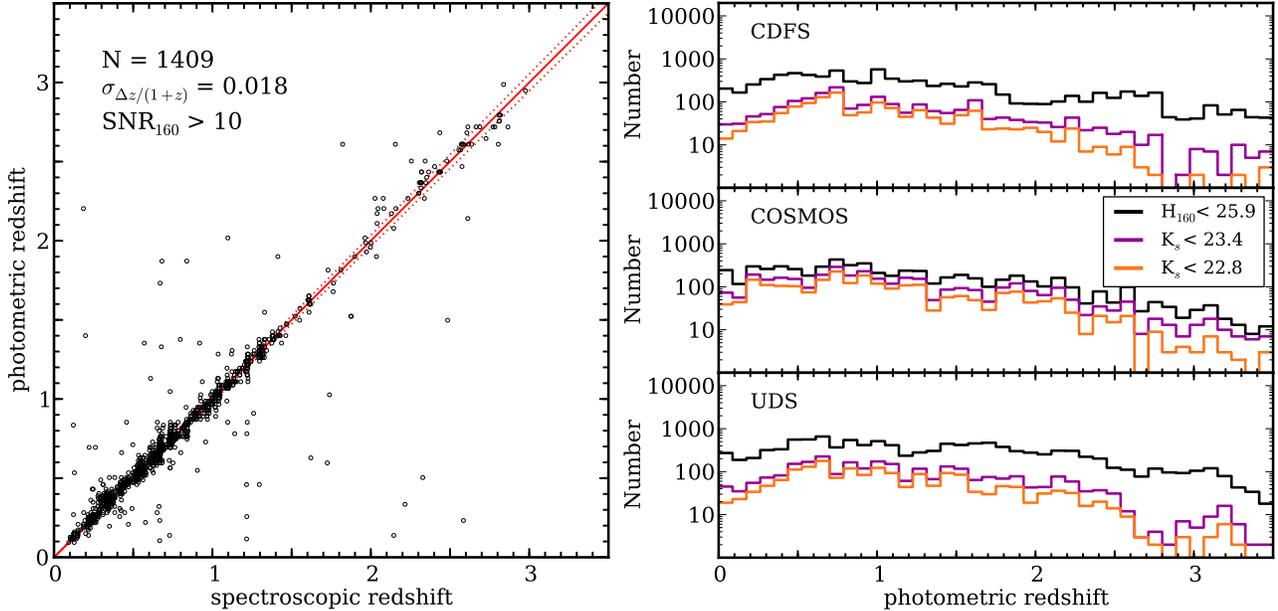 , width=0.95\linewidth }
\caption{\figtxt 
  $left$: Comparison of spectroscopic to photometric \z s
  across all three pointings of \zf . Only secure spectroscopic \z s
  of objects at SNR$_{160}  > 10$ are considered.
We find a NMAD scatter of 0.018 in $\Delta z / ( 1 + z )$ shown by the
red dotted lines with about 3\% of sources being catastrophic outliers
($|\Delta z / ( 1 + z_{spec} ) | > 0.15$). 
  $right$: Redshift
  distributions in each \zf\ field corresponding to our estimated
  magnitude limit (black) as well as the magnitude limits of
  UltraVISTA \citep[purple;][]{McCracken12} and NMBS
  \citep[orange;][]{Brammer11}. 
}
\label{zdist}
\end{figure*}

In this paper we extend measurements of the SMF to masses $\gtrsim$ 1
dex deeper than results from recent large surveys at $0.2 < z < 3$. 
Studies over the past decade have revealed that the luminosity
function and the SMF are not well-characterized by a standard \sch\
function \citep{Schechter76} due
to a steepening of the slope at stellar masses below $10^{10}$\msun\
\citep[e.g.][]{Baldry04,Blanton05}. Beyond $z \sim 1$ no survey has been deep and wide enough to
accurately constrain the low-mass end of the SMF. 
Here we use new data from the FourStar Galaxy Evolution Survey (\zf)
to construct the deepest measurement of the SMF to date.  We find a
visible upturn in the total SMF at $< 10^{10}$\msun\ as early as $z =
2$.  Furthermore, we measure the SMF of star-forming and \qui\
galaxies separately, finding that these populations evolve differently
with cosmic time. The star-forming SMF grows slowly, while the
quiescent SMF grows much more rapidly, especially at low masses; we
find that the \qui\ fraction of $9 <$ \logm\ $< 10$ galaxies 
increases by $\sim$5\x\ from $z \approx 2$ to $z \approx 0.1$ indicating that a large number of low-mass
star-forming galaxies are becoming quenched.
In this work we take into account uncertainties due to photometric
redshifts, stellar mass estimates, the classification of galaxies at
star-forming versus \qui\, and cosmic variance. 


All magnitudes are in the absolute bolometric system (AB).
We denote magnitudes measured in the HST WFC3 F125W and F160W filters
as $J_{125}$ and $H_{160}$ respectively.  The symbol \mstar\ is
reserved for the characteristic mass of the \sch\ function, and we
assume a $\Lambda$CDM cosmology throughout with \Om = 0.3 , \Olam =
0.7 and $h$ = 0.7.

%
%

\section{Data and Analysis}


\subsection{Photometry}

We make use of the deep near-IR imaging from the FourStar Galaxy
Evolution survey (\zf; Straatman et al. in prep) conducted using the FourStar
imager \citep{Persson13} on the 6.5m Magellan Baade telescope at Las Campanas
Observatory.
The use of medium-band filters in the near-IR \citep{vanDokkum09}
allows us to accurately sample wavelengths bracketing the Balmer
break of galaxies leading to more well-constrained photometric \z s at
$1<z<4$ than with broadband filters alone. 
In conjunction with existing optical through mid-IR photometry, 
this dataset provides a comprehensive sampling of the
0.3 -- 8\um\ spectral energy distribution (SED) of galaxies.

The ZFOURGE data and photometry are described in detail by Straatman
et al. (in prep.). Here we provide a brief summary.
\zf\ is composed of three $11' \times 11'$ pointings with coverage in
the CDFS \citep{Giacconi02}, COSMOS \citep{Capak07} and UDS
\citep{Lawrence07}.
The 5$\sigma$ depth in a circular aperture of 0.6'' diameter in $K_s$ is
24.8, 25.2 and 24.6 in the CDFS, COSMOS and UDS fields respectively.
Typical seeing was 0.8" or better in the ground based bands.
All optical-NIR images were convolved to a moffat PSF with
FWHM=0.75''; for some images this meant deconvolution from an
originally larger PSF. Fluxes were then measured within a circular
aperture of 0.8". 
Since image quality is much lower in the $Spitzer$ IRAC bands 
this photometry was first deblended using the $H_{160}$ image with the
techniques of \citet{Labbe06}.
Apart from the $Spitzer$ IRAC imaging, blending and source confusion
is a minor issue.

The \zf\ fields also benefit from HST imaging taken as part of the
CANDELS survey \citep{Grogin11,Koekemoer11}. We utilize the $J_{125}$ and $H_{160}$ imaging which
reach \twid 26.5 mag, significantly deeper than our ground-based
medium-band data. The high S/N photometry aids in photometric redshift
estimates even though the filters are broader than our ground-based
medium-band data. We also use the $H_{160}$ data as our detection
image. But because some of the faintest sources in the $H_{160}$
images are not detected in our ground-based data -- and thus will have
poorly constrained SEDs -- we limit our study to objects detected at
SNR$_{160} > 10$ (corresponding to $H_{160} \sim 25.9$) as a threshold
to remove galaxies that are poorly detected at other wavelengths.
In the Appendix we show examples of galaxies near our adopted flux
limit; as can be seen, these galaxies are strongly-detected and have
well-constrained photometric redshifts.
The total area of our final sample with full coverage in \zf\ and
CANDELS is \twid 316 arcmin$^2$.

We also make use of data from the NEWFIRM Medium-Band Survey
\citep[NMBS;][]{Whitaker11} which includes imaging
in the same set of medium-band near-IR filters as \zf . 
The similarity of the photometry helps reduce any inter-survey systematics.
NMBS is composed of two $30' \times 30'$ pointings in the AEGIS
\citep{Davis07} and COSMOS \citep{Capak07} fields. 
The COSMOS pointing encompasses one of our ZFOURGE fields; in the
region of overlap we make use of the higher-quality \zf\ data as
opposed to the NMBS data.
Photometric \z s from NMBS are shown to have a scatter of $\sigma_z /
(1 + z)$ = 0.017, 0.008 in the AEGIS and COSMOS fields respectively
when compared to spectroscopic \z s \citep{Whitaker11}.
Although NMBS is shallower than the rest of our sample, 
reaching depths of \twid 24.5 mag in $J_1$, $J_2$, $J_3$ and \twid
23.5 mag in $H_s$, $H_l$, $K_s$,
including it increases our survey area by a factor of 5.3 allowing us
to much better constrain the high-mass end of the SMF
\citep[see][]{Brammer11}.

\subsection{Photometric Redshifts \& Stellar Masses}

We use the public SED-fitting code {\tt EAZY} \citep{Brammer08} to measure
photometric \z s and rest-frame colors. {\tt EAZY} utilizes a default
set of 6 spectral templates that include prescriptions for emission
lines derived from the {\tt PEGASE} models \citep{Fioc97} 
plus an additional dust-reddened template derived from the
\cite{Maraston05} models. 
Linear combinations of these templates are fit to the 0.3 -- 8\um\
photometry for each galaxy to estimate \z s.

Figure \ref{zdist} demonstrates the accuracy of our photometric
\z s in comparison to available spectroscopic \z s. Only sources
reported with secure spectroscopic detections are considered.
Overall, we find a normalized median absolute deviation (NMAD) scatter
of 1.8\% in $\Delta z / ( 1 + z_{spec} )$. 
At $z<1.5$ this scatter becomes 1.7\% with about 2.7\% of catastrophic
failures ($ |\Delta z / ( 1 + z_{spec} ) | > 0.15$).
As we push to $z>1.5$, where the balmer break of galaxies \z s into
the medium-band NIR filters, this scatter becomes 2.2\% with about 9\%
of catastrophic failures. 
We note here that this scatter is likely biased upward
since objects with secure spectroscopic \z s tend to be strongly
star-forming systems with weak balmer breaks, and thus do not
benefit the most from the deep medium-band NIR photometry from \zf .
Also shown in Figure \ref{zdist} are \z\ distributions in the three
fields of \zf\ corresponding to our estimated magnitude limit as well
as the magnitude limits of UltraVISTA \citep{McCracken12}
and NMBS \citep{Brammer11} in black, purple and orange respectively.
 Spectroscopic \z s from CDFS come from 
\citet{Vanzella08}, 
\citet{LeFevre05}, 
\citet{Szokoly04}, 
\citet{Doherty05}, 
\citet{Popesso09}, and
\citet{Balestra10}.
Spectroscopic \z s from UDS come from 
\citet{Simpson12} and
\citet{Smail08}.
Spectroscopic \z s from COSMOS come from the
NASA/IPAC Infrared Science Archive\footnotemark[9].
\footnotetext[9]{ http://irsa.ipac.caltech.edu/data/COSMOS/ }

However a comparison to spectroscopic samples can be of limited use,
since the result of such comparisons depend strongly on how the
spectroscopic objects were selected. Moreover, fainter objects and
more distant objects, which are more difficult to detect
spectroscopically, are also expected to have larger photometric
redshift errors. We use the close-pairs analysis of \citet{Quadri10}
to estimate the typical uncertainties for the full sample of 
objects in our catalog, finding $\sigma \approx 0.02$ at $z \sim 0.5$
and this increases to $\sigma \approx 0.05$ at $z \sim 2.5$.

To obtain stellar masses we use the {\tt FAST} code \citep{Kriek09} which
fits stellar population synthesis models to the measured SEDs of
galaxies to infer various galactic properties.
Specifically, we use models from \citet{Bruzual03} following an
exponentially declining star-formation history assuming a
\citet{Chabrier03} initial mass function. 
We assume solar metallicity and allow A$_v$ to vary
between $[0, 4]$. 

We note here that stellar masses derived from SED-fitting
are dependent on assumed parameters in the models (metallicity, dust
law, stellar population models etc.). Variations in these assumptions
have been shown to lead to systematic offsets in stellar masses as
opposed to random errors
\citep[e.g.][]{Maraston05,Marchesini09,Conroy09}, however a full
investigation of these effects is beyond the scope of this paper.

\begin{figure}[t]
\epsfig{ file=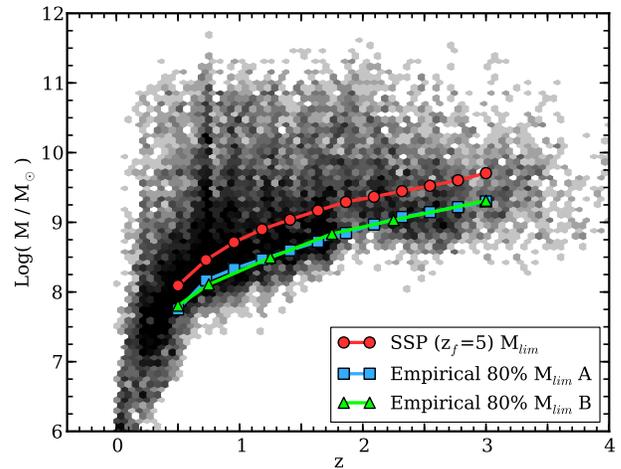 , width=0.95\linewidth }
\caption{\figtxt Galaxy stellar mass as a function of \z\ for our
  $H_{160}$-selected sample.
Our empirically derived 80\% mass-completeness limits from
down-scaling galaxies to our SNR limit and from magnitude-mass
diagrams are shown in blue and green respectively (see section
\ref{masscompleteness}).
Both techniques yield nearly identical limits. 
Also shown is the completeness limit determined from passively
evolving a SSP with a formation \z\ $z_f = 5$ which we adopt as a
separate mass-completeness limit for the \qui\ population.
}
\label{masslims}
\end{figure}

\subsection{Stellar Mass Completeness}
\label{masscompleteness}

Understanding the mass-completeness limits of our dataset is crucial
to our analysis.
\citet{Marchesini09} describe a technique whereby a sample of galaxies
below the nominal flux-completeness limit is taken from a deeper survey.
These galaxies were then scaled up in flux and mass to the completeness
limit of their survey. The resulting distribution in mass forms a
representative sample of the most massive galaxies that 
could just escape detection in their sample. 
The upper envelope of this distribution will therefore represent an
empirical determination of the \z -dependent mass-completeness limit.

In the absence of deeper data, \citet{Quadri12} modified this
technique slightly by using a sample that lies above the flux
completeness limit and scaling the fluxes and the masses down; this is
the method adopted here.
We start with all galaxies that are a factor 2 - 3\x\ above our
signal-to-noise threshold (SNR$_{160} > 10$) and scale down their
masses by the appropriate factor. 
From this scaled down sample we take the upper envelope
that encompasses 80\% of the galaxies as the \z -dependent
mass-completeness limit, shown in blue in Figure \ref{masslims}.

To obtain another measurement of the mass-completeness limit, we
employ a similar technique to that described by \citet{Chang13}.  First we
estimate the magnitude limit corresponding to our SNR threshold to be
$H_{160} \approx 25.9$. Then, in narrow bins of mass we calculate the
fraction of galaxies that are brighter than this magnitude at all SNRs.
At the
highest stellar masses this is 100\% but gradually decreases as we
probe towards lower masses. We search for the mass-bin where this
fraction is 80\% at various \z s, which we take as the
mass-completeness limit. The results from this technique are shown in
green in Figure \ref{masslims}.

Both of the empirical techniques above are performed on all galaxies
(i.e. without distinguishing star-forming/\qui ) and yield nearly
identical values which gives us confidence in our
measurements. 
We use these mass-completeness limits for both the total and
star-forming SMFs. 

\begin{figure}[t]
\epsfig{ file=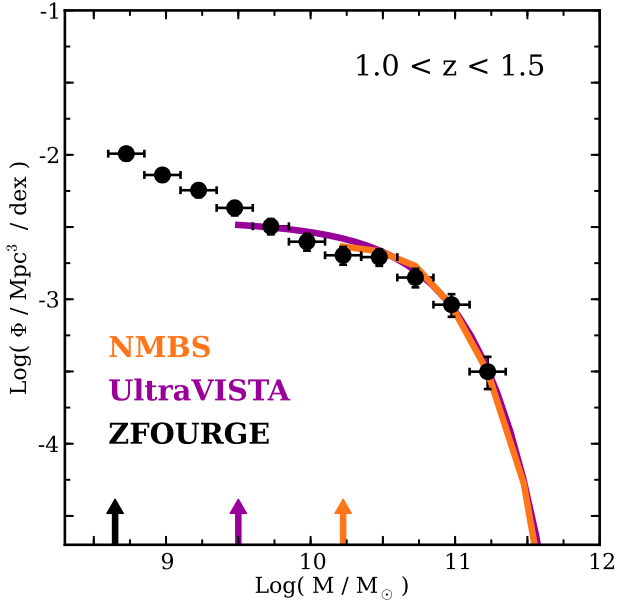 , width=0.95\linewidth }
\caption{\figtxt The stellar mass function at $z = [1.0, 1.5]$ as measured
by recent deep surveys compared to \zf : NMBS
\citep[orange;][]{Brammer11} and UltraVISTA
\citep[purple;][]{Muzzin13}. 
Arrows indicate the respective mass-completeness limits.
Errorbars shown here represent total 1$\sigma$ errors as described in
section \ref{unc}.
Previous studies were not able to reach a great enough depth over a
significant area to reveal the steepening of the SMF. }
\label{depths}
\end{figure}


Finally, since \qui\ galaxies have higher mass-to-light ratios than
the general population, the corresponding mass-completeness limit will
be higher. We calculate this mass-completeness limit from
a stellar population synthesis model obtained from EzGal
\citep{Mancone12}.  Specifically, we consider a single stellar
population (SSP) following a \citet{Chabrier03} IMF, of solar
metallicity formed at a \z\ of $z_f = 5$.
The mass-completeness limit derived from this approach is
representative of the oldest galaxies at a given \z\ since $z_f$. 
We adopt this as the mass-completeness limit for \qui\
galaxies, shown in red in Figure \ref{masslims}.

\begin{figure}[t]
\epsfig{ file=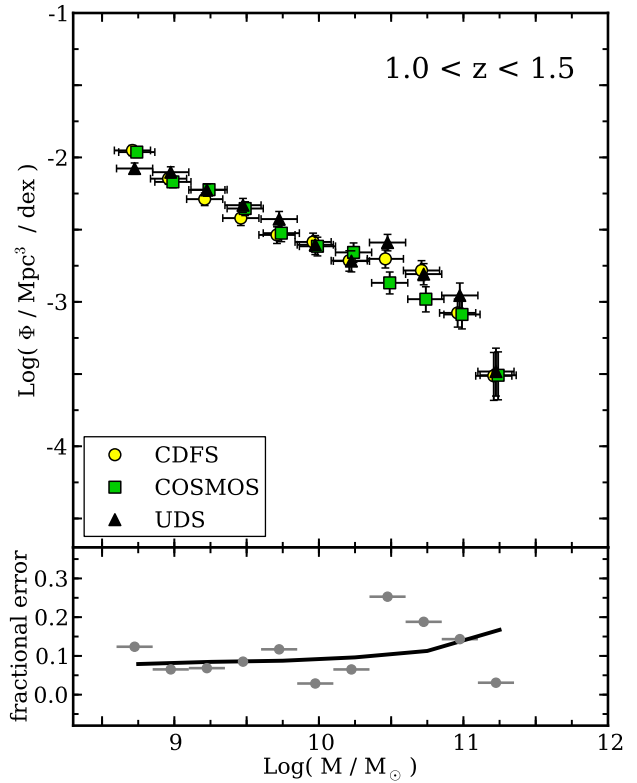 , width=0.95\linewidth }
\caption{\figtxt The stellar mass function at $z = [1.0, 1.5]$
  measured independently in all three \zf\ subfields (excluding data
  from NMBS).
Our total combined survey area is \twid 316 arcmin$^2$.
Errorbars shown here represent Poisson and SED-fitting uncertainties
but exclude cosmic variance estimates.
In the bottom panel we show the fractional uncertainty introduced by cosmic
variance determined from the standard deviation in the SMF among the
three fields (gray points).
The black line shows the predicted uncertainty using prescriptions
from \cite{Moster11} which is in agreement with the scatter we see
among our independent SMFs.
  }
\label{cvariance}
\end{figure}

\begin{figure*}[t]
\epsfig{ file=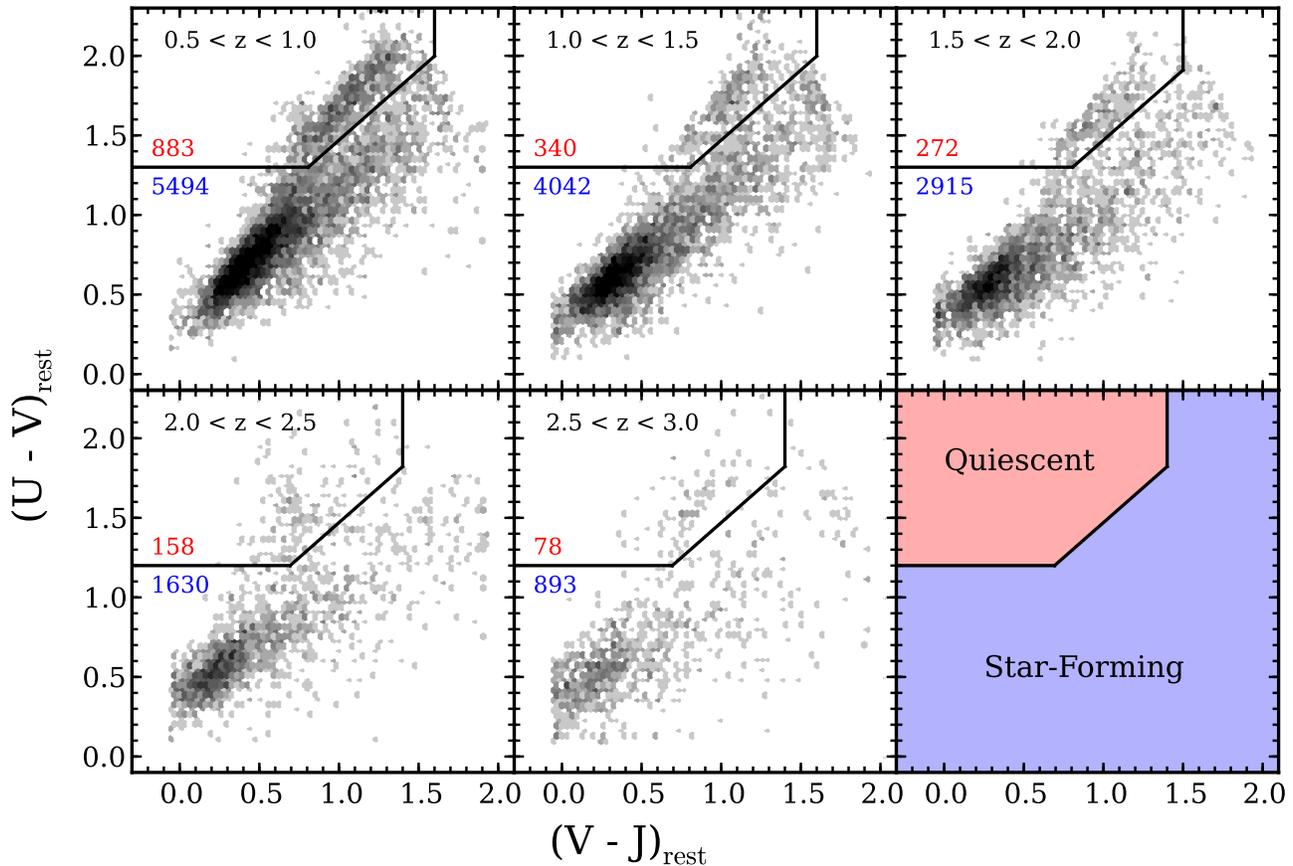, width=0.95\linewidth }
\caption{ Rest-frame $UVJ$ diagrams used to separate star-forming and
  \qui\ galaxies as indicated in the bottom-right panel. Only galaxies
  above our mass-completeness limits are shown  from 
  all three \zf\ pointings (CDFS, COSMOS, UDS; the NMBS data are excluded).
  In each panel, the number of \qui\ and star-forming galaxies are
  shown in the selection regions in red and blue respectively.
  Due to the similarity between our dataset and NMBS, we use the
  \z -dependent
 selection regions (shown in black) determined in \cite{Whitaker11}.
}
\label{uvjfigs}
\end{figure*}

Our data provide a view of the SMF to depths that have previously been
inaccessible over significant areas. In Figure \ref{depths} we plot
an example SMF as measured by \zf , UltraVISTA \citep{Muzzin13} and
NMBS \citep{Brammer11} which reach $K_s$-band 5$\sigma$ depths of
about 24.9, 23.4 and 22.8 magnitudes respectively.
Furthermore, since \zf\ is split into three independent pointings,
errors due to cosmic variance are suppressed compared to a survey
of equal area composed of one pointing. We show an example of
field-to-field variance in Figure \ref{cvariance} where we plot the
SMF measured from each \zf\ pointing individually.

\subsection{Selection of Star-Forming \& \Qui\ Galaxies}

In this work we divide the full galaxy sample into star-forming and \qui\
populations. We separate these populations in a rest-frame $U-V$
vs. $V-J$ color-color diagram (hereafter $UVJ$ diagram), which has
been shown to effectively trace the galaxy color-bimodality as far as
$z=3$ \citep{Labbe05,Williams09,Patel12,Whitaker11,Muzzin13}. 
The strength of this technique lies in its weak dependence on dust
extinction, since the dust-reddening vector tends not to scatter
galaxies across the selection boundary.
This helps avoid contamination by dusty
star-forming galaxies in typical red-sequence selection techniques.

We derive rest-frame $U-V$ and $V-J$ colors from the best-fit {\tt EAZY}
templates to the observed photometry. 
In Figure \ref{uvjfigs} we show $UVJ$ diagrams for our galaxy sample at
various \z s. Only galaxies above their respective
mass-completeness limit are shown. 
The bimodality can be see to $z \sim 3$.


\subsection{Uncertainties}
\label{unc}

The accuracy with which we are able to measure the SMF
is dependent on multiple steps, each having its own 
uncertainty. 
Poisson uncertainties ($\sigma_{poisson}$) are calculated using
prescriptions from \citet{Gehrels86}. We also include cosmic
variance ($\sigma_{cv}$) and uncertainties in the SED
modeling used to estimate photometric \z s, rest-frame colors and
stellar masses ($\sigma_{sed}$).

We calculate cosmic variance as a function of \z\ and mass using the
{\tt getcv} routine described in \citet{Moster11}. This yields 
cosmic variance uncertainties that range from $\approx$25\% at
$10^{11}$\msun\ to $\approx$8\% at $10^{8.5}$\msun . Cosmic variance
can also be estimated from the scatter in 
the SMFs of the independent pointings from \zf\ (Figure
\ref{cvariance}). Overall, we find this scatter to be consistent with
the predictions. 


To estimate the uncertainty contribution from SED modeling we conduct 100 Monte
Carlo simulations on our catalogs. For each realization we
independently perturb photometric \z s and stellar masses using the
68\% confidence limits output from {\tt EAZY} and {\tt FAST}.
SMFs are then recalculated over the same \z\ ranges used
throughout. The 1$\sigma$ scatter in the resulting SMFs is then taken
as the \z - and mass-dependent uncertainty. These uncertainties
range from 5 - 15\% over the span of \z s in this study.

\begin{figure*}[t]
\epsfig{ file=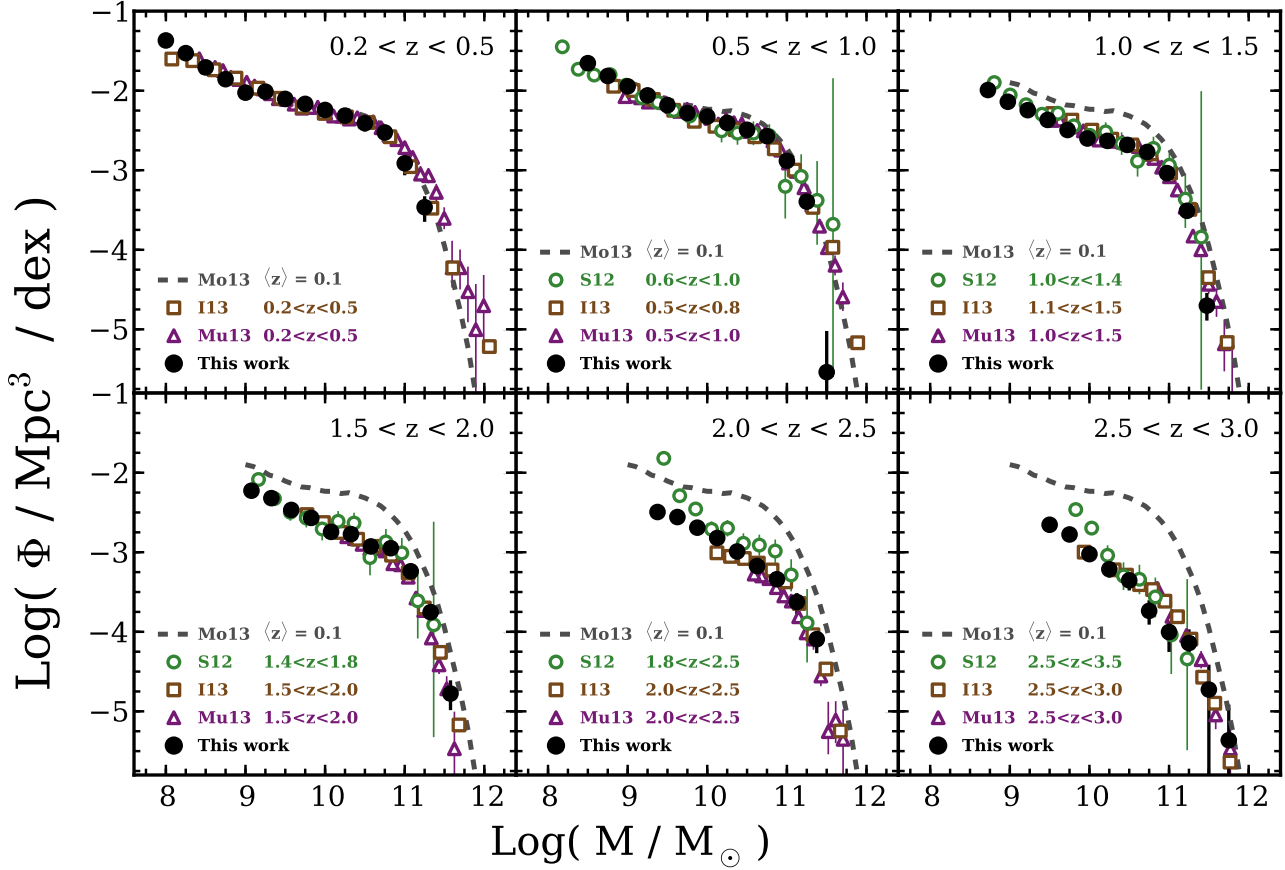, width=0.95\linewidth }
\caption{ \figtxt
Stellar mass functions for all galaxies between $0.2 < z < 3$ with
errorbars representing total 1$\sigma$ uncertainties. 
We compare our SMFs to those from other recent studies:
\citet{Moustakas13} (Mo13),
\citet{Santini12} (S12),
\citet{Ilbert13} (I13), and
\citet{Muzzin13} (Mu13).
Data are only shown above the reported mass-completeness limit for
each study. 
There is excellent agreement where the SMFs overlap except with the
$z>2$ SMF from \citet{Santini12}.
}
\label{mfcompare}
\end{figure*}

\begin{figure*}
\epsfig{ file=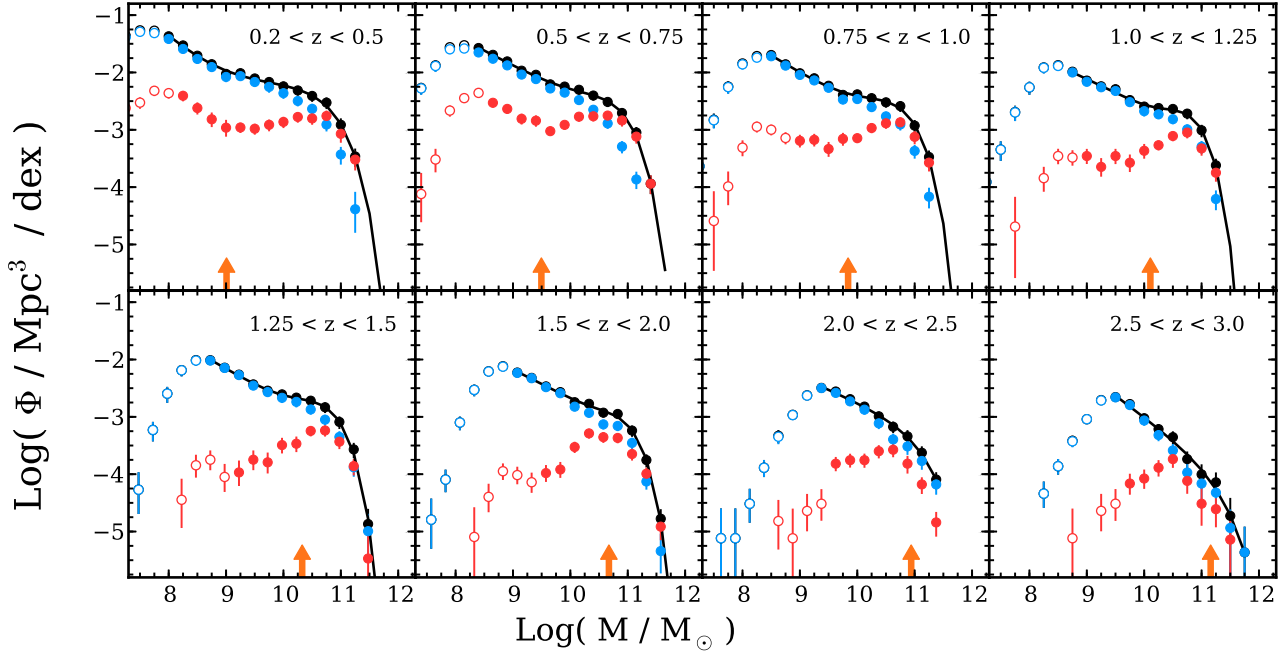, width=0.95\linewidth }
\caption{ \figtxt Stellar mass functions in sequential \z\ bins for all
  ($black$), star-forming ($blue$) and \qui\ ($red$) galaxies.
Open symbols correspond to data below each subsample's respective
mass-completeness limit.
We have used data from NMBS to supplement the high-mass end of
each SMF down to the limits indicated by the orange arrows.
Best-fit \sch\ functions to the total SMF are plotted as black lines.
Even as far as $z \sim 2$ the total SMF exhibits a low-mass upturn.
Furthermore, we show a clear decline in the \qui\ SMF below \mstar\
towards high-$z$, which cannot be attributed to incompleteness.
}
\label{mfs}
\end{figure*}

\begin{figure*}[t]
\epsfig{ file=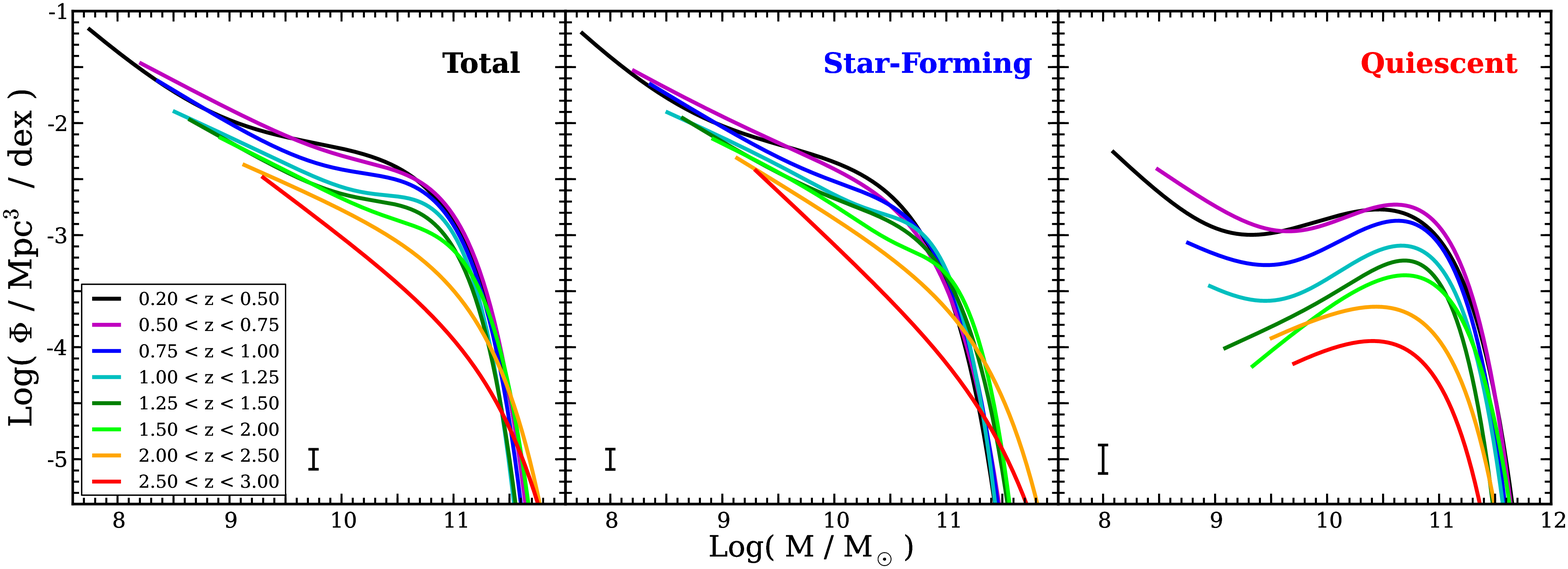, width=0.95\linewidth }
\caption{ \figtxt
Evolution of our total ($left$), star-forming ($middle$) and \qui\
($right$) SMFs between $0.2 < z < 3$. For each \z\ bin we only plot
where we are above the corresponding mass-completeness
limit. Errorbars in the lower-left of each panel show representative
1$\sigma$ uncertainties that include Poisson errors, cosmic variance
and SED-fitting uncertainties. Double-\sch\ fits are used at $z < 2$
for the total and star-forming SMFs and at $z < 1.5$ for the \qui\
SMF. 
}
\label{summary}
\end{figure*}

\begin{figure*}[p]
\epsfig{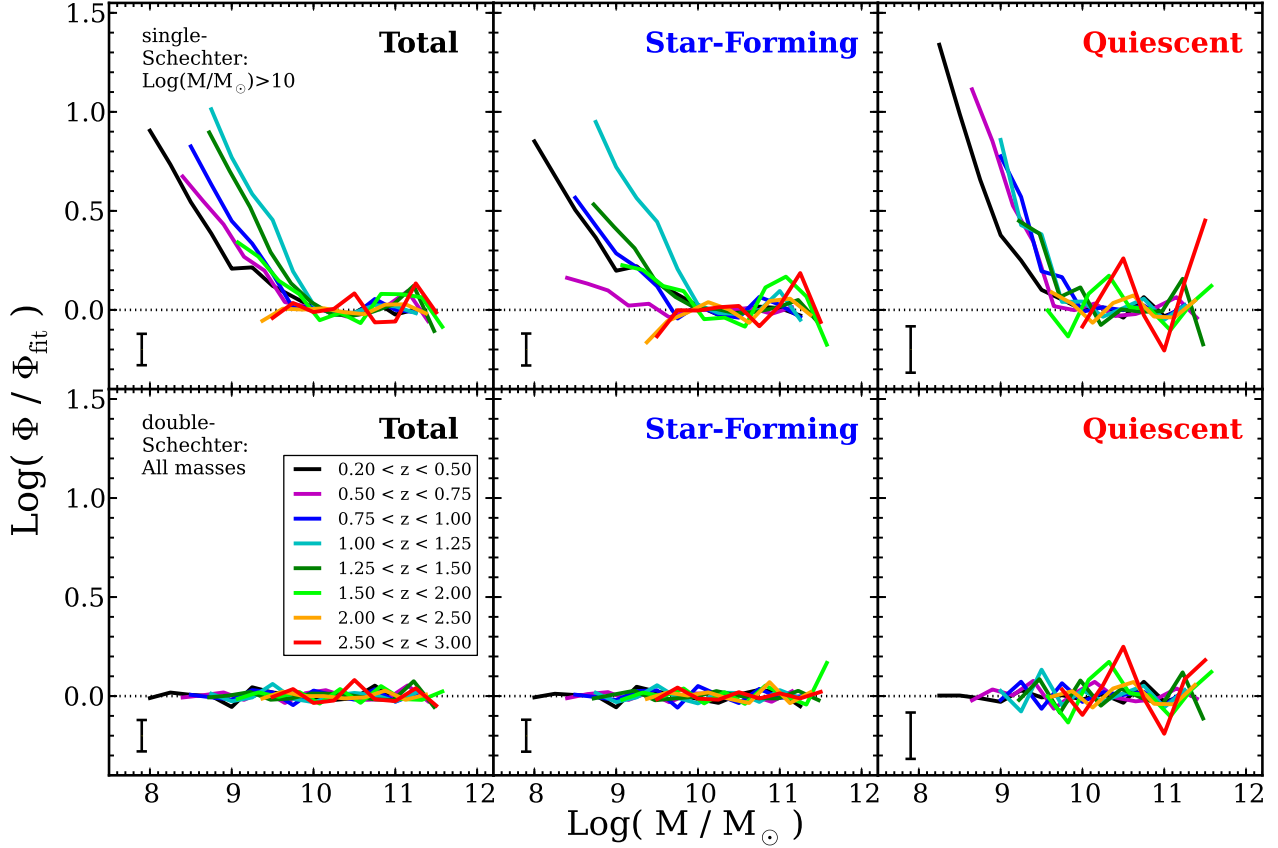}
\caption{ \figtxt
Residuals from functional fits to our total, star-forming and \qui\
SMFs. Errorbars in the lower-left of each panel show representative
1$\sigma$ uncertainties.
The top three panels correspond to single-\sch\ functions fit at
Log(M/\msun ) $> 10$. Residuals here clearly show the presence of the
low-mass upturn at $z<2$ in the total and star-forming SMFs and at
$z<1.5$ in the \qui\ SMF. 
The bottom three panels correspond to double-\sch\ functions fit at
all masses. Residuals here are consistent with random noise,
indicating that the double-\sch\ function is an accurate description
of the SMF.
However, although the double-\sch\ function provides a good fit, a
single-\sch\ function is sufficient for our SMFs at $z>2$. 
}
\label{residuals}
\end{figure*}


We consider another source of uncertainty involved in the
classification of galaxies as star-forming vs. \qui .
Here, we evaluate the statistical uncertainty associated with the $UVJ$
classification based on photometric uncertainties ($\sigma_{uvj}$).
To do this we perform 100 Monte Carlo simulations on a sample of
galaxies at $10 <$ SNR$_{160} < 200$, perturbing fluxes according to a
Gaussian probability density function based on 1$\sigma$ photometric
uncertainties. Photometric \z s and rest-frame colors were remeasured
for each iteration, from which galaxies were reclassified as being
star-forming or \qui . We find that at fixed SNR$_{160}$ more galaxies
scatter into vs. out of the \qui\ region, boosting the \qui\
fraction.
However this effect is small; we find that the quiescent fractions
typically vary by $<2\%$, and that this value has a scatter of
$<0.4\%$ between the simulations. 
A particular concern may be that the number density of quiescent
sources at low masses may be significantly affected by a small
fraction of the (much more abundant) star-forming galaxies scattering
into the quiescent region, but we find that this is not a major
concern.

In total, our uncertainty budgets become:

\begin{equation}
\begin{split}
\sigma_{\rm tot} = & \sqrt{ \sigma^2_{poisson} + \sigma^2_{cv} +  \sigma^2_{sed} } \\
\sigma_{\rm sf/qui} = & \sqrt{ \sigma^2_{poisson} + \sigma^2_{cv} +  \sigma^2_{sed} + \sigma^2_{uvj} }
\end{split}
\end{equation}

\section{Results}

\subsection{Measuring the Stellar Mass Function}

In Figure \ref{mfcompare} we show our measurements of the total SMF
over $0.2 < z < 3$.
For comparison we have included corresponding measurements at
similar redshifts intervals from recent works
\citep{Santini12,Moustakas13,Ilbert13,Muzzin13}.
We find excellent agreement in the regions of overlap,
except with \citet{Santini12} who measure higher densities of
  galaxies at $z>2$.

In Figure \ref{mfs} we subdivide the total SMF into star-forming and \qui\
populations over the same range of \z s as in Figure
\ref{mfcompare}. The data are also presented in Table 1. We reiterate that these mass functions have been
supplemented  by NMBS to provide better constraints at the
high-mass end. Orange arrows show the mass-limits for the contribution
of NMBS to each SMF in Figure \ref{mfs}. 
We also show the growth of each SMF (total, star-forming and \qui ) in
Figure \ref{summary} over our entire \z\ range.

In calculating the SMF, we include only galaxies that lie above the
mass-completeness limit corresponding to the upper \z\ limit of each 
subsample. 
We follow the procedures outlined in \citet{Avni80} to combine the
multiple fields of our survey in calculating SMFs.
The SMF ($\Phi$) is then simply calculated as:

\begin{equation}
\Phi (M)  =  \;  \frac{1}{\Delta M}  \;  \sum_{i=1}^N \frac{1}{V_c}  
\end{equation}

\noindent
where
$M = $ \logm , 
$\Delta M$ is the size of the mass-bin,
$N$ is the number of galaxies in the mass-bin between the \z\
limits ($z_{\mathrm{min}}$, $z_{\mathrm{max}}$)
and $V_c$ is the comoving volume based on the survey area and \z\
limits.
We refrain from using the 1/\vmax\ formalism
\citep{Avni80} to avoid introducing any potential bias associated with
evolution in the SMF over our relatively wide \z\ bins.
Since we do not apply a 1/\vmax\ correction, $V_c$ is the
same for all galaxies in a given \z\ bin.

%
%

\begin{figure*}[t]
\epsfig{ file=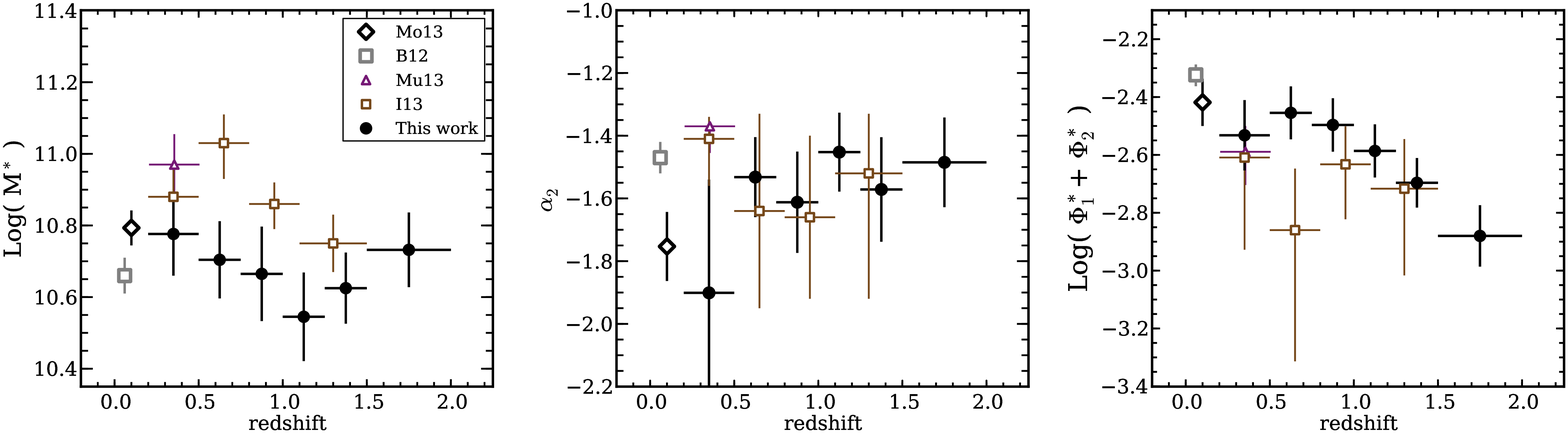, width=0.95\linewidth }
\caption{ \figtxt
Redshift evolution in the best-fit \sch\ parameters for total SMFs at
all \z s where a double-\sch\ function provides a better fit.
$left$: Best-fit values for the characteristic mass $M^*$
$center$: Best-fit values for the low-mass slope $\alpha_2$.
$right$: Sum of the best-fit values for the normalizations
($\Phi^*_1$, $\Phi^*_2$). 
For comparison we also show measurements from other studies that 
found the double-\sch\ to provide a better fit:
\citet{Moustakas13} (Mo13),
\citet{Baldry12} (B12),
\citet{Muzzin13} (Mu13),
\citet{Ilbert13} (I13). 
We note that we fit a double-\sch\ function to the SMF from
\citet{Moustakas13} ourselves as no such parameters were reported.
The parameters we assume are given in section \ref{smfevo}.
The only statistically significant evolution we find in our data is in
Log($\Phi^*_1 + \Phi^*_2$) indicating that the shape of the SMF
remains mostly constant but increases in normilization with time.
 }
\label{alphamstarphistar}
\end{figure*}

\begin{table*}
\label{smfstable}
\begin{center}
\caption{Stellar Mass Functions}

 Total \\[0.3mm]
\begin{tabular}{c|cccccccc}

\hline \\[-2.3mm]  
\hline \\[-2.1mm]  
 &   $0.2 < z < 0.5$  &  $0.5 < z < 0.75$  &  $0.75 < z < 1.0$  &  $1.0 < z < 1.25$  &  $1.25 < z < 1.5$  &  $1.5 < z < 2.0$  &  $2.0 < z < 2.5$  &  $2.5 < z < 3.0$ \\[1mm] 
Log(M/\msun )  & Log($\Phi$) & Log($\Phi$) & Log($\Phi$) & Log($\Phi$) & Log($\Phi$) & Log($\Phi$)  & Log($\Phi$)  & Log($\Phi$) \\[1mm] 
\hline \\[-2mm]

8.00  &  $-1.37^{+0.06}_{-0.07}$  &  ---  &  ---  &  ---  &  ---  &  ---  &  ---  &  --- \\
8.25  &  $-1.53^{+0.06}_{-0.07}$  &  $-1.53^{+0.06}_{-0.07}$  &  ---  &  ---  &  ---  &  ---  &  ---  &  --- \\
8.50  &  $-1.71^{+0.07}_{-0.08}$  &  $-1.60^{+0.05}_{-0.06}$  &  $-1.70^{+0.05}_{-0.06}$  &  ---  &  ---  &  ---  &  ---  &  --- \\
8.75  &  $-1.86^{+0.07}_{-0.08}$  &  $-1.76^{+0.06}_{-0.06}$  &  $-1.86^{+0.05}_{-0.06}$  &  $-1.99^{+0.06}_{-0.06}$  &  $-2.02^{+0.06}_{-0.07}$  &  ---  &  ---  &  --- \\
9.00  &  $-2.03^{+0.08}_{-0.09}$  &  $-1.86^{+0.06}_{-0.07}$  &  $-2.01^{+0.06}_{-0.06}$  &  $-2.14^{+0.06}_{-0.07}$  &  $-2.14^{+0.06}_{-0.07}$  &  $-2.20^{+0.05}_{-0.06}$  &  ---  &  --- \\
9.25  &  $-2.01^{+0.07}_{-0.08}$  &  $-2.00^{+0.06}_{-0.07}$  &  $-2.10^{+0.06}_{-0.07}$  &  $-2.24^{+0.06}_{-0.07}$  &  $-2.28^{+0.06}_{-0.07}$  &  $-2.31^{+0.05}_{-0.06}$  &  $-2.53^{+0.06}_{-0.07}$  &  --- \\
9.50  &  $-2.10^{+0.07}_{-0.09}$  &  $-2.12^{+0.07}_{-0.08}$  &  $-2.23^{+0.06}_{-0.07}$  &  $-2.29^{+0.06}_{-0.07}$  &  $-2.46^{+0.07}_{-0.08}$  &  $-2.41^{+0.05}_{-0.06}$  &  $-2.50^{+0.06}_{-0.07}$  &  $-2.65^{+0.06}_{-0.07}$ \\
9.75  &  $-2.17^{+0.08}_{-0.10}$  &  $-2.21^{+0.06}_{-0.07}$  &  $-2.39^{+0.07}_{-0.08}$  &  $-2.48^{+0.07}_{-0.08}$  &  $-2.53^{+0.07}_{-0.08}$  &  $-2.54^{+0.06}_{-0.06}$  &  $-2.63^{+0.06}_{-0.07}$  &  $-2.78^{+0.07}_{-0.08}$ \\
10.00  &  $-2.24^{+0.08}_{-0.10}$  &  $-2.25^{+0.06}_{-0.08}$  &  $-2.45^{+0.07}_{-0.09}$  &  $-2.59^{+0.08}_{-0.09}$  &  $-2.61^{+0.08}_{-0.09}$  &  $-2.67^{+0.06}_{-0.07}$  &  $-2.74^{+0.07}_{-0.08}$  &  $-3.02^{+0.08}_{-0.09}$ \\
10.25  &  $-2.31^{+0.08}_{-0.09}$  &  $-2.35^{+0.07}_{-0.08}$  &  $-2.45^{+0.07}_{-0.09}$  &  $-2.73^{+0.08}_{-0.10}$  &  $-2.68^{+0.08}_{-0.09}$  &  $-2.76^{+0.06}_{-0.07}$  &  $-2.91^{+0.08}_{-0.09}$  &  $-3.21^{+0.09}_{-0.10}$ \\
10.50  &  $-2.41^{+0.08}_{-0.10}$  &  $-2.45^{+0.07}_{-0.09}$  &  $-2.52^{+0.08}_{-0.09}$  &  $-2.64^{+0.07}_{-0.09}$  &  $-2.71^{+0.08}_{-0.09}$  &  $-2.87^{+0.07}_{-0.08}$  &  $-3.07^{+0.09}_{-0.10}$  &  $-3.35^{+0.10}_{-0.13}$ \\
10.75  &  $-2.53^{+0.09}_{-0.11}$  &  $-2.55^{+0.08}_{-0.09}$  &  $-2.59^{+0.08}_{-0.10}$  &  $-2.72^{+0.08}_{-0.10}$  &  $-2.84^{+0.08}_{-0.10}$  &  $-3.03^{+0.08}_{-0.09}$  &  $-3.35^{+0.10}_{-0.13}$  &  $-3.74^{+0.13}_{-0.17}$ \\
11.00  &  $-2.91^{+0.11}_{-0.15}$  &  $-2.82^{+0.09}_{-0.11}$  &  $-2.93^{+0.10}_{-0.13}$  &  $-3.01^{+0.10}_{-0.12}$  &  $-3.12^{+0.10}_{-0.13}$  &  $-3.13^{+0.08}_{-0.10}$  &  $-3.54^{+0.12}_{-0.16}$  &  $-4.00^{+0.18}_{-0.25}$ \\
11.25  &  $-3.46^{+0.14}_{-0.18}$  &  $-3.32^{+0.10}_{-0.13}$  &  $-3.47^{+0.11}_{-0.15}$  &  $-3.62^{+0.11}_{-0.15}$  &  $-3.65^{+0.12}_{-0.16}$  &  $-3.56^{+0.10}_{-0.13}$  &  $-3.89^{+0.12}_{-0.17}$  &  $-4.14^{+0.17}_{-0.28}$ \\
11.50  &  ---  &  ---  &  ---  &  ---  &  $-4.99^{+0.30}_{-0.41}$  &  $-4.27^{+0.12}_{-0.15}$  &  $-4.41^{+0.14}_{-0.19}$  &  $-4.73^{+0.31}_{-2.00}$ \\[1mm] \hline
\end{tabular}
\\[1.5mm]
 Star-Forming \\[0.3mm]
\begin{tabular}{c|cccccccc}

\hline \\[-2.3mm]  
\hline \\[-2.1mm]  
 &   $0.2 < z < 0.5$  &  $0.5 < z < 0.75$  &  $0.75 < z < 1.0$  &  $1.0 < z < 1.25$  &  $1.25 < z < 1.5$  &  $1.5 < z < 2.0$  &  $2.0 < z < 2.5$  &  $2.5 < z < 3.0$ \\[1mm] 
Log(M/\msun )  & Log($\Phi$) & Log($\Phi$) & Log($\Phi$) & Log($\Phi$) & Log($\Phi$) & Log($\Phi$)  & Log($\Phi$)  & Log($\Phi$) \\[1mm] 
\hline \\[-2mm]

8.00  &  $-1.42^{+0.06}_{-0.07}$  &  ---  &  ---  &  ---  &  ---  &  ---  &  ---  &  --- \\
8.25  &  $-1.59^{+0.06}_{-0.07}$  &  $-1.60^{+0.06}_{-0.07}$  &  ---  &  ---  &  ---  &  ---  &  ---  &  --- \\
8.50  &  $-1.76^{+0.07}_{-0.08}$  &  $-1.67^{+0.05}_{-0.06}$  &  $-1.72^{+0.05}_{-0.06}$  &  ---  &  ---  &  ---  &  ---  &  --- \\
8.75  &  $-1.91^{+0.07}_{-0.08}$  &  $-1.83^{+0.06}_{-0.06}$  &  $-1.88^{+0.05}_{-0.06}$  &  $-2.00^{+0.06}_{-0.06}$  &  $-2.03^{+0.06}_{-0.07}$  &  ---  &  ---  &  --- \\
9.00  &  $-2.08^{+0.08}_{-0.09}$  &  $-1.92^{+0.06}_{-0.07}$  &  $-2.04^{+0.06}_{-0.06}$  &  $-2.16^{+0.06}_{-0.07}$  &  $-2.15^{+0.06}_{-0.07}$  &  $-2.20^{+0.05}_{-0.06}$  &  ---  &  --- \\
9.25  &  $-2.06^{+0.07}_{-0.08}$  &  $-2.09^{+0.06}_{-0.07}$  &  $-2.14^{+0.06}_{-0.07}$  &  $-2.26^{+0.06}_{-0.07}$  &  $-2.29^{+0.06}_{-0.07}$  &  $-2.32^{+0.05}_{-0.06}$  &  $-2.53^{+0.06}_{-0.07}$  &  --- \\
9.50  &  $-2.17^{+0.07}_{-0.09}$  &  $-2.19^{+0.07}_{-0.08}$  &  $-2.27^{+0.06}_{-0.07}$  &  $-2.32^{+0.06}_{-0.07}$  &  $-2.48^{+0.07}_{-0.08}$  &  $-2.42^{+0.05}_{-0.06}$  &  $-2.51^{+0.06}_{-0.07}$  &  $-2.66^{+0.06}_{-0.07}$ \\
9.75  &  $-2.25^{+0.08}_{-0.10}$  &  $-2.28^{+0.06}_{-0.07}$  &  $-2.47^{+0.07}_{-0.08}$  &  $-2.52^{+0.07}_{-0.08}$  &  $-2.55^{+0.07}_{-0.08}$  &  $-2.56^{+0.06}_{-0.06}$  &  $-2.67^{+0.06}_{-0.07}$  &  $-2.79^{+0.07}_{-0.08}$ \\
10.00  &  $-2.36^{+0.08}_{-0.10}$  &  $-2.39^{+0.07}_{-0.08}$  &  $-2.55^{+0.08}_{-0.09}$  &  $-2.68^{+0.08}_{-0.09}$  &  $-2.68^{+0.08}_{-0.09}$  &  $-2.73^{+0.06}_{-0.07}$  &  $-2.78^{+0.07}_{-0.08}$  &  $-3.06^{+0.08}_{-0.09}$ \\
10.25  &  $-2.50^{+0.08}_{-0.09}$  &  $-2.55^{+0.07}_{-0.08}$  &  $-2.60^{+0.07}_{-0.09}$  &  $-2.88^{+0.09}_{-0.10}$  &  $-2.75^{+0.08}_{-0.10}$  &  $-2.89^{+0.07}_{-0.07}$  &  $-3.00^{+0.08}_{-0.09}$  &  $-3.32^{+0.09}_{-0.11}$ \\
10.50  &  $-2.63^{+0.09}_{-0.11}$  &  $-2.76^{+0.08}_{-0.09}$  &  $-2.77^{+0.08}_{-0.09}$  &  $-2.81^{+0.07}_{-0.09}$  &  $-2.87^{+0.08}_{-0.09}$  &  $-3.07^{+0.07}_{-0.09}$  &  $-3.26^{+0.09}_{-0.11}$  &  $-3.59^{+0.11}_{-0.14}$ \\
10.75  &  $-2.91^{+0.10}_{-0.12}$  &  $-3.00^{+0.08}_{-0.10}$  &  $-2.91^{+0.09}_{-0.11}$  &  $-2.99^{+0.08}_{-0.10}$  &  $-3.07^{+0.08}_{-0.10}$  &  $-3.26^{+0.09}_{-0.10}$  &  $-3.54^{+0.11}_{-0.14}$  &  $-3.97^{+0.16}_{-0.20}$ \\
11.00  &  $-3.43^{+0.13}_{-0.18}$  &  $-3.46^{+0.10}_{-0.13}$  &  $-3.37^{+0.10}_{-0.13}$  &  $-3.29^{+0.10}_{-0.13}$  &  $-3.39^{+0.10}_{-0.13}$  &  $-3.35^{+0.09}_{-0.11}$  &  $-3.69^{+0.13}_{-0.17}$  &  $-4.16^{+0.20}_{-0.28}$ \\
11.25  &  $-4.39^{+0.30}_{-0.41}$  &  $-4.30^{+0.20}_{-0.25}$  &  $-4.17^{+0.16}_{-0.20}$  &  $-4.21^{+0.15}_{-0.20}$  &  $-3.95^{+0.13}_{-0.17}$  &  $-3.85^{+0.10}_{-0.13}$  &  $-4.00^{+0.13}_{-0.17}$  &  $-4.32^{+0.18}_{-0.29}$ \\
11.50  &  ---  &  ---  &  ---  &  ---  &  $-5.17^{+0.37}_{-0.52}$  &  $-4.78^{+0.17}_{-0.21}$  &  $-4.59^{+0.15}_{-0.21}$  &  $-4.94^{+0.32}_{-2.00}$ \\ \hline
\end{tabular}
\\[1.5mm]
 Quiescent \\[0.3mm]
\begin{tabular}{c|cccccccc}

\hline \\[-2.3mm]  
\hline \\[-2.1mm]  
 &   $0.2 < z < 0.5$  &  $0.5 < z < 0.75$  &  $0.75 < z < 1.0$  &  $1.0 < z < 1.25$  &  $1.25 < z < 1.5$  &  $1.5 < z < 2.0$  &  $2.0 < z < 2.5$  &  $2.5 < z < 3.0$ \\[1mm] 
Log(M/\msun )  & Log($\Phi$) & Log($\Phi$) & Log($\Phi$) & Log($\Phi$) & Log($\Phi$) & Log($\Phi$)  & Log($\Phi$)  & Log($\Phi$) \\[1mm] 
\hline \\[-2mm]

8.25  &  $-2.41^{+0.08}_{-0.10}$  &  ---  &  ---  &  ---  &  ---  &  ---  &  ---  &  --- \\
8.50  &  $-2.62^{+0.10}_{-0.11}$  &  $-2.42^{+0.07}_{-0.08}$  &  ---  &  ---  &  ---  &  ---  &  ---  &  --- \\
8.75  &  $-2.82^{+0.12}_{-0.14}$  &  $-2.58^{+0.07}_{-0.08}$  &  ---  &  ---  &  ---  &  ---  &  ---  &  --- \\
9.00  &  $-2.96^{+0.14}_{-0.16}$  &  $-2.77^{+0.09}_{-0.10}$  &  $-3.19^{+0.11}_{-0.12}$  &  $-3.46^{+0.12}_{-0.14}$  &  ---  &  ---  &  ---  &  --- \\
9.25  &  $-2.96^{+0.08}_{-0.10}$  &  $-2.75^{+0.09}_{-0.10}$  &  $-3.17^{+0.10}_{-0.12}$  &  $-3.65^{+0.15}_{-0.17}$  &  $-3.97^{+0.21}_{-0.24}$  &  ---  &  ---  &  --- \\
9.50  &  $-2.98^{+0.09}_{-0.10}$  &  $-2.94^{+0.10}_{-0.11}$  &  $-3.33^{+0.12}_{-0.14}$  &  $-3.46^{+0.13}_{-0.14}$  &  $-3.79^{+0.17}_{-0.19}$  &  $-4.14^{+0.17}_{-0.19}$  &  ---  &  --- \\
9.75  &  $-2.91^{+0.09}_{-0.11}$  &  $-2.99^{+0.07}_{-0.08}$  &  $-3.16^{+0.11}_{-0.12}$  &  $-3.57^{+0.14}_{-0.16}$  &  $-3.75^{+0.16}_{-0.18}$  &  $-3.95^{+0.14}_{-0.15}$  &  $-3.72^{+0.11}_{-0.12}$  &  $-4.16^{+0.17}_{-0.20}$ \\
10.00  &  $-2.86^{+0.09}_{-0.11}$  &  $-2.83^{+0.07}_{-0.08}$  &  $-3.16^{+0.11}_{-0.12}$  &  $-3.37^{+0.12}_{-0.14}$  &  $-3.45^{+0.12}_{-0.14}$  &  $-3.55^{+0.09}_{-0.11}$  &  $-3.76^{+0.11}_{-0.13}$  &  $-4.08^{+0.16}_{-0.18}$ \\
10.25  &  $-2.78^{+0.08}_{-0.10}$  &  $-2.78^{+0.07}_{-0.09}$  &  $-2.97^{+0.08}_{-0.09}$  &  $-3.26^{+0.11}_{-0.13}$  &  $-3.52^{+0.13}_{-0.15}$  &  $-3.35^{+0.08}_{-0.09}$  &  $-3.64^{+0.11}_{-0.12}$  &  $-3.89^{+0.13}_{-0.15}$ \\
10.50  &  $-2.80^{+0.09}_{-0.11}$  &  $-2.75^{+0.08}_{-0.09}$  &  $-2.89^{+0.08}_{-0.10}$  &  $-3.11^{+0.08}_{-0.09}$  &  $-3.24^{+0.08}_{-0.10}$  &  $-3.30^{+0.08}_{-0.09}$  &  $-3.53^{+0.10}_{-0.12}$  &  $-3.74^{+0.12}_{-0.15}$ \\
10.75  &  $-2.76^{+0.09}_{-0.12}$  &  $-2.75^{+0.08}_{-0.10}$  &  $-2.87^{+0.09}_{-0.11}$  &  $-3.05^{+0.08}_{-0.10}$  &  $-3.23^{+0.09}_{-0.11}$  &  $-3.40^{+0.09}_{-0.11}$  &  $-3.82^{+0.13}_{-0.16}$  &  $-4.12^{+0.18}_{-0.22}$ \\
11.00  &  $-3.07^{+0.12}_{-0.16}$  &  $-2.93^{+0.09}_{-0.11}$  &  $-3.12^{+0.10}_{-0.13}$  &  $-3.33^{+0.10}_{-0.13}$  &  $-3.46^{+0.10}_{-0.13}$  &  $-3.54^{+0.09}_{-0.11}$  &  $-4.08^{+0.17}_{-0.22}$  &  $-4.51^{+0.27}_{-0.38}$ \\
11.25  &  $-3.52^{+0.14}_{-0.19}$  &  $-3.37^{+0.11}_{-0.14}$  &  $-3.57^{+0.12}_{-0.15}$  &  $-3.75^{+0.12}_{-0.16}$  &  $-3.95^{+0.13}_{-0.17}$  &  $-3.87^{+0.10}_{-0.13}$  &  $-4.54^{+0.15}_{-0.21}$  &  $-4.61^{+0.19}_{-0.32}$ \\
11.50  &  ---  &  ---  &  ---  &  ---  &  $-5.47^{+0.52}_{-0.90}$  &  $-4.44^{+0.13}_{-0.16}$  &  $-4.89^{+0.19}_{-0.26}$  &  $-5.14^{+0.34}_{-2.00}$ \\ \hline

\end{tabular}
\end{center}

\end{table*}

\begin{table*}
\label{bestfitstable2x}
\begin{center}
\caption{Best-fit double-\sch\ Parameters}
 Total \\[0.3mm]
\begin{tabular}{c|ccccccc}

\hline \\[-2.3mm]  
\hline \\[-2.1mm]  
Redshift  & Log($M^*$)   &   $\alpha_1$ &
Log($\Phi^*_1$)   &  $\alpha_2$ &  Log($\Phi^*_2$) &
$\chi^2_{\mathrm{red}}$ \\[1mm] 
\hline \\[-2mm]

$0.20 < z < 0.50$ & $10.78 \pm0.11$ & $-0.98 \pm0.24$ & $-2.54 \pm0.12$ & $-1.90 \pm0.36$ & $-4.29 \pm0.55$ & 0.3 \\
$0.50 < z < 0.75$ & $10.70 \pm0.10$ & $-0.39 \pm0.50$ & $-2.55 \pm0.09$ & $-1.53 \pm0.12$ & $-3.15 \pm0.23$ & 0.5 \\
$0.75 < z < 1.00$ & $10.66 \pm0.13$ & $-0.37 \pm0.49$ & $-2.56 \pm0.09$ & $-1.61 \pm0.16$ & $-3.39 \pm0.28$ & 0.6 \\
$1.00 < z < 1.25$ & $10.54 \pm0.12$ & $0.30 \pm0.65$ & $-2.72 \pm0.10$ & $-1.45 \pm0.12$ & $-3.17 \pm0.19$ & 0.8 \\
$1.25 < z < 1.50$ & $10.61 \pm0.08$ & $-0.12 \pm0.49$ & $-2.78 \pm0.08$ & $-1.56 \pm0.16$ & $-3.43 \pm0.23$ & 0.3 \\
$1.50 < z < 2.00$ & $10.74 \pm0.09$ & $0.04 \pm0.62$ & $-3.05 \pm0.11$ & $-1.49 \pm0.14$ & $-3.38 \pm0.20$ & 0.8 \\
$2.00 < z < 2.50$ & $10.69 \pm0.29$ & $1.03 \pm1.64$ & $-3.80 \pm0.30$ & $-1.33 \pm0.18$ & $-3.26 \pm0.23$ & 0.4 \\
$2.50 < z < 3.00$ & $10.74 \pm0.31$ & $1.62 \pm1.88$ & $-4.54 \pm0.41$ & $-1.57 \pm0.20$ & $-3.69 \pm0.28$ & 1.3 \\ \hline

\end{tabular}
\\[1.5mm]
Star-Forming \\
\begin{tabular}{c|ccccccc}

\hline \\[-2.3mm]
\hline \\[-2.1mm]
Redshift  & Log($M^*$)   &   $\alpha_1$ &  Log($\Phi^*_1$)   &  $\alpha_2$ &  Log($\Phi^*_2$) &  $\chi^2_{\mathrm{red}}$ \\[1mm]
\hline \\[-2mm]

$0.20 < z < 0.50$ & $10.59 \pm0.09$ & $-1.08 \pm0.23$ & $-2.67 \pm0.11$ & $-2.00 \pm0.49$ & $-4.46 \pm0.63$ & 0.3 \\
$0.50 < z < 0.75$ & $10.65 \pm0.23$ & $-0.97 \pm1.32$ & $-2.97 \pm0.28$ & $-1.58 \pm0.54$ & $-3.34 \pm0.67$ & 0.3 \\
$0.75 < z < 1.00$ & $10.56 \pm0.13$ & $-0.46 \pm0.63$ & $-2.81 \pm0.10$ & $-1.61 \pm0.17$ & $-3.36 \pm0.28$ & 0.9 \\
$1.00 < z < 1.25$ & $10.44 \pm0.11$ & $0.53 \pm0.73$ & $-2.98 \pm0.14$ & $-1.44 \pm0.11$ & $-3.11 \pm0.16$ & 0.8 \\
$1.25 < z < 1.50$ & $10.69 \pm0.12$ & $-0.55 \pm0.69$ & $-3.04 \pm0.12$ & $-1.62 \pm0.24$ & $-3.59 \pm0.35$ & 0.2 \\
$1.50 < z < 2.00$ & $10.59 \pm0.10$ & $0.75 \pm0.70$ & $-3.37 \pm0.16$ & $-1.47 \pm0.10$ & $-3.28 \pm0.13$ & 0.9 \\
$2.00 < z < 2.50$ & $10.58 \pm0.18$ & $2.06 \pm1.43$ & $-4.30 \pm0.39$ & $-1.38 \pm0.15$ & $-3.28 \pm0.18$ & 0.7 \\
$2.50 < z < 3.00$ & $10.61 \pm0.22$ & $2.36 \pm1.84$ & $-4.95 \pm0.49$ & $-1.67 \pm0.19$ & $-3.71 \pm0.25$ & 0.8 \\ \hline

\end{tabular}
\\[1.5mm]
Quiescent \\
\begin{tabular}{c|ccccccc}

\hline \\[-2.3mm]
\hline \\[-2.1mm]
Redshift  & Log($M^*$)   &   $\alpha_1$ &  Log($\Phi^*_1$)   &  $\alpha_2$ &  Log($\Phi^*_2$) &  $\chi^2_{\mathrm{red}}$ \\[1mm]
\hline \\[-2mm]

$0.20 < z < 0.50$ & $10.75 \pm0.10$ & $-0.47 \pm0.20$ & $-2.76 \pm0.09$ & $-1.97 \pm0.34$ & $-5.21 \pm0.48$ & 0.2 \\
$0.50 < z < 0.75$ & $10.68 \pm0.07$ & $-0.10 \pm0.27$ & $-2.67 \pm0.05$ & $-1.69 \pm0.24$ & $-4.29 \pm0.33$ & 0.9 \\
$0.75 < z < 1.00$ & $10.63 \pm0.12$ & $0.04 \pm0.44$ & $-2.81 \pm0.05$ & $-1.51 \pm0.67$ & $-4.40 \pm0.56$ & 0.4 \\
$1.00 < z < 1.25$ & $10.63 \pm0.12$ & $0.11 \pm0.44$ & $-3.03 \pm0.05$ & $-1.57 \pm0.81$ & $-4.80 \pm0.61$ & 0.8 \\
$1.25 < z < 1.50$ & $10.49 \pm0.11$ & $0.85 \pm1.07$ & $-3.36 \pm0.30$ & $-0.54 \pm0.66$ & $-3.72 \pm0.44$ & 0.6 \\
$1.50 < z < 2.00$ & $10.77 \pm0.18$ & $-0.19 \pm0.96$ & $-3.41 \pm0.23$ & $-0.18 \pm1.21$ & $-3.91 \pm0.51$ & 1.9 \\
$2.00 < z < 2.50$ & $10.69 \pm0.14$ & $-0.37 \pm0.52$ & $-3.59 \pm0.10$ & $-3.07 \pm16.13$ & $-6.95 \pm1.66$ & 0.7 \\
$2.50 < z < 3.00$ & $9.95  \pm0.23$ & $-0.62  \pm2.63$ & $-4.22 \pm0.41$ & $2.51 \pm2.43$ & $-4.51  \pm0.62$ & 1.7 \\ \hline

\end{tabular}
\end{center}

$^a$  In units of \msun \\
$^b$  In units of Mpc$^{-3}$

\end{table*}

\begin{table}
\label{bestfitstable1x}
\begin{center}
\caption{Best-fit single-\sch\ Parameters}
 Total \\[0.3mm]
\begin{tabular}{c|ccccccc}

\hline \\[-2.3mm]
\hline \\[-2.1mm]
Redshift  & Log($M^*$)$^a$   &   $\alpha$ &  Log($\Phi^*$)$^b$   &  $\chi^2_{\mathrm{red}}$ \\[1mm]
\hline \\[-2mm]

$0.20 < z < 0.50$ & $11.05 \pm0.10$ & $-1.35 \pm0.04$ & $-2.96 \pm0.10$ & 3.3 \\
$0.50 < z < 0.75$ & $11.00 \pm0.06$ & $-1.35 \pm0.04$ & $-2.93 \pm0.07$ & 4.6 \\
$0.75 < z < 1.00$ & $11.16 \pm0.12$ & $-1.38 \pm0.04$ & $-3.17 \pm0.11$ & 4.5 \\
$1.00 < z < 1.25$ & $11.09 \pm0.10$ & $-1.33 \pm0.05$ & $-3.19 \pm0.11$ & 4.2 \\
$1.25 < z < 1.50$ & $10.88 \pm0.05$ & $-1.29 \pm0.05$ & $-3.11 \pm0.08$ & 5.6 \\
$1.50 < z < 2.00$ & $11.03 \pm0.05$ & $-1.33 \pm0.05$ & $-3.28 \pm0.08$ & 4.5 \\
$2.00 < z < 2.50$ & $11.13 \pm0.13$ & $-1.43 \pm0.08$ & $-3.59 \pm0.14$ & 0.3 \\
$2.50 < z < 3.00$ & $11.35 \pm0.33$ & $-1.74 \pm0.12$ & $-4.36 \pm0.29$ & 1.0 \\ \hline

\end{tabular}
\\[1.5mm]
Star-Forming \\
\begin{tabular}{c|ccccccc}

\hline \\[-2.3mm]
\hline \\[-2.1mm]
Redshift  & Log($M^*$)   &   $\alpha$ &  Log($\Phi^*$)   &  $\chi^2_{\mathrm{red}}$ \\[1mm]
\hline \\[-2mm]

$0.20 < z < 0.50$ & $10.73 \pm0.06$ & $-1.37 \pm0.04$ & $-2.94 \pm0.08$ & 2.4 \\
$0.50 < z < 0.75$ & $10.79 \pm0.07$ & $-1.42 \pm0.04$ & $-3.04 \pm0.08$ & 0.7 \\
$0.75 < z < 1.00$ & $10.86 \pm0.07$ & $-1.43 \pm0.04$ & $-3.16 \pm0.09$ & 3.0 \\
$1.00 < z < 1.25$ & $10.85 \pm0.07$ & $-1.37 \pm0.05$ & $-3.20 \pm0.09$ & 3.2 \\
$1.25 < z < 1.50$ & $10.89 \pm0.05$ & $-1.38 \pm0.05$ & $-3.27 \pm0.08$ & 2.1 \\
$1.50 < z < 2.00$ & $10.97 \pm0.05$ & $-1.45 \pm0.05$ & $-3.44 \pm0.08$ & 4.2 \\
$2.00 < z < 2.50$ & $11.28 \pm0.19$ & $-1.60 \pm0.08$ & $-3.96 \pm0.19$ & 1.3 \\
$2.50 < z < 3.00$ & $11.49 \pm0.46$ & $-1.93 \pm0.12$ & $-4.82 \pm0.38$ & 1.4 \\ \hline

\end{tabular}
\\[1.5mm]
Quiescent \\
\begin{tabular}{c|ccccccc}

\hline \\[-2.3mm]
\hline \\[-2.1mm]
Redshift  & Log($M^*$)   &   $\alpha$ &  Log($\Phi^*$)   &  $\chi^2_{\mathrm{red}}$ \\[1mm]
\hline \\[-2mm]

$0.20 < z < 0.50$ & $11.11 \pm0.14$ & $-0.98 \pm0.07$ & $-3.18 \pm0.10$ & 4.1 \\
$0.50 < z < 0.75$ & $11.03 \pm0.08$ & $-0.98 \pm0.07$ & $-3.15 \pm0.09$ & 9.9 \\
$0.75 < z < 1.00$ & $10.88 \pm0.09$ & $-0.59 \pm0.10$ & $-3.00 \pm0.08$ & 2.1 \\
$1.00 < z < 1.25$ & $10.84 \pm0.09$ & $-0.47 \pm0.11$ & $-3.16 \pm0.07$ & 2.1 \\
$1.25 < z < 1.50$ & $10.60 \pm0.04$ & $-0.03 \pm0.14$ & $-3.17 \pm0.05$ & 0.9 \\
$1.50 < z < 2.00$ & $10.76 \pm0.05$ & $-0.14 \pm0.12$ & $-3.29 \pm0.05$ & 1.8 \\
$2.00 < z < 2.50$ & $10.73 \pm0.08$ & $-0.49 \pm0.18$ & $-3.63 \pm0.09$ & 0.4 \\
$2.50 < z < 3.00$ & $10.65 \pm0.19$ & $-0.43 \pm0.34$ & $-3.92 \pm0.14$ & 1.6 \\ \hline

\end{tabular}
\end{center}

$^a$  In units of \msun \\
$^b$  In units of Mpc$^{-3}$

\end{table}

\subsection{Fitting the Stellar Mass Function}

The depth of our survey allows us to test for the shape of the SMF,
namely we fit both single- and double-\sch\ functions to determine
the best fit. The single-\citet{Schechter76} function is defined as:

\begin{equation}
\Phi (M) dM = \: \mathrm{ln} (10) \; \Phi^* \left[ 10^{(M-M^*)(1+\alpha)}
\right] \mathrm{exp} ( -10^{(M-M^*)} ) dM
\end{equation}

\noindent
where again $M = $ \logm , $\alpha$ is the slope of the power-law at
low masses, \phistar\ is the normalization and \mstar\ is the
characteristic mass. The double-\sch\ function is defined as:

\begin{equation}
\begin{split}
\Phi (M) dM = \;\; &\Phi_1 (M) dM + \Phi_2 (M) dM  \\
= \;\; &\mathrm{ln} (10) \; \mathrm{exp} \left( -10^{(M-M^*)} \right) 10^{(M-M^*)} \\
&\times \left[ \Phi^*_1 10^{(M-M^*) \alpha_1}   +   \Phi^*_2 10^{(M-M^*) \alpha_2}   \right] dM
\end{split}
\end{equation}

\noindent
where again $M = $ \logm , ($\alpha_1$, $\alpha_2$) are the slopes and ($\Phi^*_1$,
$\Phi^*_2$) are the normalizations of the constituent \sch\
functions respectively,  and \mstar\ again is the characteristic
mass. Note that one value for \mstar\ is used for both constituents in
the double-\sch\ function. This functional form of the
double-\sch\ function is the same as in \citet{Baldry08}.


Recent measurements of the total SMF at $z<1.5$ have shown that the
SMF steepens at M<$10^{10}$\msun\
\citep[e.g.][]{Baldry08,Li09,Drory09,Pozzetti10,Moustakas13,Ilbert13,Muzzin13}. 
We fit each of our mass functions with both single- and double-\sch\
functions. 
We show best-fit parameters as well as reduced chi-squared values for
each in tables 2 and 3. 
From the reduced chi-squared values we find that the total SMF is much
better fit by a double-\sch\ function at $z \leq 2$.
At $z > 2$ we find that a single-\sch\ function is sufficient,
however this may be because we do not go deep enough to detect
significant structure at low masses.

This is clearly shown in Figure \ref{residuals} where we
plot the residuals of both single- and double-\sch\ fits to the
total SMF.
A prominent upturn is revealed in the top three panels where we fit
single-\sch\ functions at \logm\ $>10$ only. 
In the bottom three panels we show the residuals from fitting
double-\sch\ functions at all masses, which are consistent within our
measurement uncertainties.
However, although the double-\sch\ provides a good fit at all \z s, we
find that a single-\sch\ works just as well at $z>1.5$ for the \qui\
SMF and at $z>2$ for the total and star-forming SMFs.
We observe the same behavior even if the NMBS data is excluded from
the calculation, proving that the steepening of the low-mass slope is
not caused by a systematic offset between the surveys we use.
In fact there is evidence for a steepening in each of the three \zf\
fields independently.



\subsection{The Weakly-Evolving Shape of the Total Stellar Mass Function}
\label{smfevo}

The left panel of Figure \ref{alphamstarphistar} shows the best-fit
values for \mstar\ as a function of \z .
There is little statistically significant evolution in \mstar\ at $z <
2$, in agreement with other studies \citep{Marchesini09,Santini12, Muzzin13}.  
We note that our values of \mstar\ are \twid 0.2 dex lower than these
previous studies. We find that this offset is the result of comparing
single- versus double-\sch\ fits to the SMF.
The weak evolution in \mstar\ suggests that the physical
mechanism(s) responsible for the exponential cutoff in the SMF has a
mass scale that is independent of redshift \citep[see also ][]{Peng10}.

We show the best-fit values for the faint-end slope $\alpha$ as a
function of \z\ in the middle panel of Figure \ref{alphamstarphistar}.
We plot only the steeper slope of ($\alpha_1$, $\alpha_2$) which
dominates at the lowest masses.
We find no statistically significant evolution in the low-mass slope
within our \z\ range.
Some evolution in alpha may be suggested when comparing to the $z \sim
0$ SMF from \citet{Moustakas13}, however we note that those authors do
not probe below $10^9$\msun , and thus do not strongly constrain the
slope at the lowest masses. We do find better agreement with the $z
\sim 0$ SMF from \citet{Baldry12}, who reach lower masses.

In the last panel of Figure \ref{alphamstarphistar} we show the redshift evolution of
$\Phi^*_1 + \Phi^*_2$. In contrast with the apparent constancy of
\mstar\ and $\alpha$, we find clear evolution in 
\phistar . Thus to rough approximation the shape of the total SMF
does not evolve over $0 < z < 2$, but the normalization does. 
\citet{Moustakas13} do not report parameters for functional fits to
their measured SMF; therefore, we fit our own doube-\sch\ function to
their $z \approx 0.1$ SMF. The best-fit parameters we find for Log(\mstar )
, $\alpha_1$, Log($\Phi^*_1$), $\alpha_2$, Log($\Phi^*_2$) are
10.79, $-$0.74, $-$2.44, $-$1.75, $-$3.69 respectively.

\subsection{Buildup of the Star-Forming and \Qui\ Populations}

\begin{figure*}[t]
\epsfig{ file=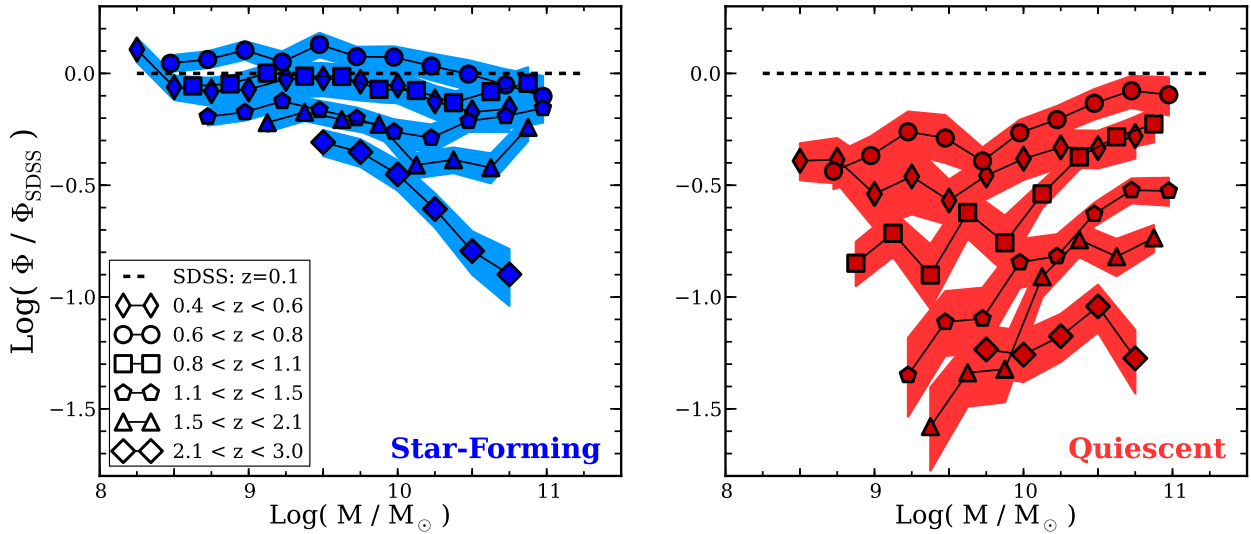, width=0.95\linewidth }
\caption{ \figtxt 
Growth in the star-forming ({\it left}) and \qui\ ({\it right}) SMFs
relative to the $z \approx 0$ star-forming and \qui\ SMFs from
\citet{Moustakas13}. Shaded regions show $1 \sigma$ Poisson and
SED-fitting uncertainties. 
Cosmic variance uncertainties are neglected for clarity, but range
between 0.05 and 0.14 dex. 
Each \z\ interval here at $z \geq 0.4$ has
been chosen to span roughly 1.2 Gyr of galaxy evolution. 
We find that the growth in the number density of star-forming galaxies
is remarkably uniform at \logm $< 10$.
The \qui\ SMF, however, exhibits a rapid increase towards lower
stellar masses. Specifically, at \logm $\leq 10$ \qui\ galaxies
increase in number by a factor of 15 - 30 whereas star-forming galaxies
increase by only a factor of 1.5 - 2. Despite the large difference in
these growth rates, star-forming galaxies still remain the dominant
population at low masses at all \z s.
 }
\label{buildup}
\end{figure*}

In Figure \ref{buildup} we show the growth in the number density
of galaxies as a function of mass in several redshift bins for the
star-forming and \qui\ subpopulations.
We show this growth by normalizing our star-forming/\qui\ SMFs to the
most recent measurements of the star-forming/\qui\ SMFs at $z \approx
0$ from SDSS \citep{Moustakas13}.
The redshift ranges at $z > 0.4$ in Figure \ref{buildup} are chosen to
track the evolution in similar time intervals of approximately 1.2 Gyr.

At \logm\ $< 11$, where we have sufficient statistics to trace the
evolution of the mass function, we find that the SMF of star forming
galaxies grows moderately with cosmic time, by 1.5 - 2.5\x\ since
$z \sim 2$. There is a hint that it actually decreases with time at $z
< 0.6$.
Only between $2<z<3$ do we observe a large jump in the number number
of star-forming galaxies at \logm $>10$.
These results are consistent with
previous works which have generally found that the star-forming SMF
evolves relatively weakly with redshift
\citep{Arnouts07,Bell07,Pozzetti10,Brammer11,Muzzin13}. 

The growth of \qui\ galaxies since $z \approx 2$ is much more rapid
than that of star-forming galaxies
\citep[e.g.][]{Arnouts07,Bell07}.
At masses greater than $10^{10}$\msun\ we find roughly a factor of 6
increase between $z = 2$ and $z = 0$ in agreement with previous studies,
however, at lower masses  there is a 15 - 30\x\ increase.
This is the first
clear detection of a decline in the low-mass \qui\ population towards
high-\z\ that is not affected by incompleteness. 
This rapid evolution causes the \qui\ fraction to increase by about a
factor of 5 for low-mass galaxies ($<10^{10}$\msun )
from $\approx$7\% at $z = 2$ to $\approx$34\% at $z = 0$.

\begin{figure*}[t]
\epsfig{ file=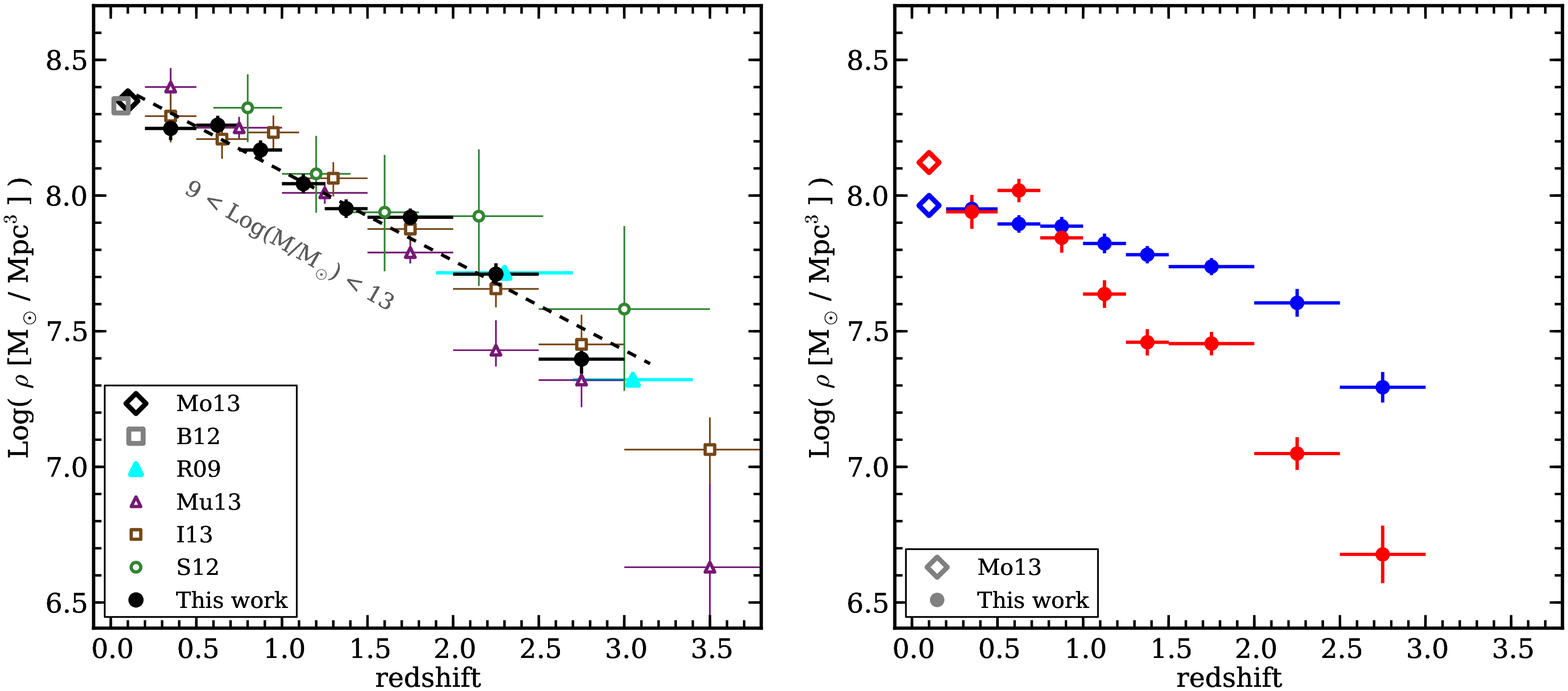, width=0.95\linewidth }
\caption{ \figtxt
Cosmic stellar mass densities as a function of \z\ evaluated from the
best-fit \sch\ functions to the total SMF ({\it left}) and the star-forming
and \qui\ SMFs ({\it right}). 
We show the total stellar mass density (integrated over $9 <$ \logm\ $<
13$) with  1$\sigma$ uncertainties determined from Monte Carlo
simulations on our SMFs. 
Other symbols show results from previous works from deep NIR surveys: 
\citet[black diamond]{Moustakas13} (Mo13), 
\citet[gray square]{Baldry12} (B12), 
\citet[purple triangles]{Muzzin13} (Mu13), 
\citet[brown squares]{Ilbert13} (I13) and
\citet[green circles]{Santini12} (S12).
The dashed black line is a least-squares fit to the ZFOURGE data:
Log$(\rho) = -0.33 (1+z) + 8.75$.
Also shown are high-\z\ mass densities inferred from a UV-selected
galaxy sample with a correction for incompleteness at low masses
\citep[cyan triangles]{Reddy09} (R09). 
Our measured mass densities are in good agreement with these previous
works.
}
\label{massdensity}
\end{figure*}

\subsection{Cosmic Stellar Mass Density}

%
%

Obtaining a precise estimate of the integrated stellar mass density in
the universe requires probing the SMF well below \mstar . Most recent
attempts at intermediate \z s have been made using near-infrared selected surveys, which make it possible to define
highly-complete samples down to some stellar mass limit. However if
this limit does not reach significantly below \mstar\ then the integrated
stellar mass density depends on an extrapolation of the observed SMF
using the best-fit \sch\ parameters \citep[e.g.][]{Marchesini09,
Santini12, Ilbert13, Muzzin13}, which
may be poorly constrained and may depend sensitively on the exact ---
and uncertain --- level of completeness near the nominal
mass-completeness limit.

In Figure \ref{massdensity} we show our measurements for the evolution
of the cosmic stellar mass densities ($\rho$) of all, star-forming and \qui\ galaxies. 
Previous studies have typically integrated best-fit \sch\
functions between $8 <$ \logm\ $< 13$, extrapolating below
mass-completeness limits wherever necessary. 
We choose to integrate our best-fits between $9 <$ \logm\ $< 13$ since
this is only marginally below our completeness limit in our highest
\z\ bin. We note here that using $10^{9}$\msun\ as opposed to
$10^{8}$\msun\ as a lower-limit decreases $\rho$ by $<5\%$.

Uncertainties are evaluated from 500 Monte Carlo simulations of the
measured SMFs. For each iteration we perturb all data points using the
combined uncertainties as described in section \ref{unc}. We then
refit \sch\ functions to recalculate \rhotot , taking the
resulting scatter as the uncertainty.
We parameterize our measurements
of the \z\ evolution of the total stellar mass density as follows:

\begin{equation}
{ \rm Log} ( \rho )   =   a (1+z) + b
\end{equation}

\noindent
where $\rho$ is the total stellar mass density in units of
\msun/Mpc$^3$. From a least-squares fit we find best-fit values of $a
= -0.33 \pm0.03$ and $b = 8.75 \pm0.07$.

Figure \ref{massdensity} also shows results from recent deep and
large-area surveys which are in overall agreement with our
measurements.
\citet{Santini12} present results using data from CANDELS Early
Release Science program in conjunction with deep ($K_s \sim 25.5$)
imaging from Hawk-I. Although their work covers significantly less
area than we present here (33 arcmin$^2$ versus 316 arcmin$^2$) our
measurements agree within $1\sigma$ uncertainties.
Measurements from the recent UltraVISTA survey \citep{McCracken12}
which covers \twid 1.6 deg$^2$ to a depth of  $K_s = 23.4$ are
presented in \citet{Ilbert13} and \citet{Muzzin13}. Our results are in
excellent agreement at all \z s except $1.5 < z < 2.5$ with
\citet{Muzzin13}.
The difference between our result and \citet{Muzzin13} is mostly due to
the large difference in the faint end slope: \citet{Muzzin13} measure a
slope of $\sim -0.9$ whereas we find $-1.33$ for the best-fit single-\sch\
function at $1.5 < z < 2.0$. \citet{Muzzin13} note that $\alpha$ is not
well constrained by their data, and do not rule out a low-mass slope
as steep as ours.

Another estimate of the stellar mass density was provided by
\citet{Reddy09}, who used an optically-selected sample of star-forming
galaxies at $1.9 < z < 3.4$ to argue that the low-mass end of the SMF is
quite steep and may have been underestimated by previous studies; they
concluded that a large fraction of the stellar mass budget of the
universe was locked up in dwarf galaxies. However these authors were
not able to probe the SMF directly given the nature of their sample
and their limited NIR and IR data, so they inferred the SMF by
performing large corrections for incompleteness. Given the depth of our
NIR-selected sample, we are able to probe down to similarly low masses
($\sim 10^{9}$\msun ) for { \it complete } samples. 

We compare our $z > 2$ measurements to estimates based on the
\citet{Reddy09} measurements in Figure \ref{massdensity}. We obtain
their value by integrating the SMF shown in their Fig. 12, and after
converting from a Salpeter IMF (N.~Reddy, private communication). The
agreement is excellent, however, as noted by those authors, their
sample is incomplete for galaxies with red colors. Thus the good
agreement that we find is partially due to the steeper slope
of their inferred SMF which is balanced by incompleteness at high
masses.\footnote{\citet{Reddy09} estimate that faint galaxies that lie
  below typical ground-based flux limits contain a roughly similar
  amount of mass as do the bright galaxies that are usually
  observed. At first glance this may seem to contradict our finding in
  Figure \ref{massdensity} that dwarf galaxies are sub-dominant. There
  are several possible explanations for this difference. One is the
  difference in the slopes of our SMFs ($\sim -1.7$ versus $\sim
  -1.4$). Another is that we limit our integration to \logm\ $> 9$,
  where we are highly complete; if we were to integrate further down the
  SMF then the contribution of dwarfs would be larger. Finally, another
  likely contributing factor is that \citet{Reddy09} select their
  sample based on the rest-frame UV emission; because of the weak
  correlation between stellar mass and UV emission, it is expected
  that UV-faint galaxies that lie below typical flux limits should
  still contain significant stellar mass.} Nonetheless, it is
encouraging that similar results are obtained using very different
types of datasets and different methods.

\section{Summary}

We have measured the galaxy stellar mass function over a broad \z\
range ($0.2 < z < 3$) utilizing data from three legacy fields with
coverage in \zf\ (CDFS, COSMOS, UDS).
We detect galaxies using deep overlapping
imaging in the $H_{160}$-band from the CANDELS survey, conducted using \emph{HST}.
This in combination with medium-band near-IR imaging from \zf\ allows us to construct a
large sample of galaxies complete to low stellar masses with accurate
photometric \z s.
Our final sample covers a combined area of 316 arcmin$^2$ to
a depth of $H_{160} = 25.9$.
We also include data from NMBS in our sample which adds \twid 1300
arcmin$^2$ at a $5\sigma$ depth of $K_s < 22.8$ to help constrain the
high-mass end.
Our data allow us to probe the SMF down to stellar masses of $\approx 
10^{9.5}$\msun\ at $z < 2.5$.

We show in Figure \ref{mfs} that the low-mass end of the \qui\ SMF
exhibits rapid evolution between $z=1.5$ and today.
We calculate greater than a factor of 10 increase in the number of
\qui\ galaxies at stellar masses $< 10^{10}$\msun .
Since the expected source of low-mass \qui\ galaxies is low-mass
star-forming galaxies that have become quenched, this leads to the
question of what is/are the dominant quenching process/processes for
low-mass galaxies.
This effect could be the result of a growing population of
low-mass galaxies being accreted onto larger halos and having their
star-formation quenched in the process.
Several studies have suggested that environmental processes become
increasingly important in the quenching of star formation at low
masses \citep[e.g.][]{Hogg03, Peng10, Geha12, Quadri12}, implying
that the differential buildup in the \qui\ SMF is at least
partially due to the evolving role of environment.

The SMF at $z \leq 1.5$ has been known to exhibit a steepening of the
faint-end slope at \logm\ $\lesssim 10$, and is thus not
well-characterized by a single-\sch\ function
\citep[e.g][]{Baldry08,Ilbert13,Muzzin13}.  We fit both single- and
double-\sch\ functions to all of our SMFs and assess which
parameterization is better based on the reduced chi-squared statistic
($\chi^2_{\mathrm{red}}$). 
Our results show that a low-mass upturn is present in the SMF up to at
least $z = 2$.
We find no evidence for evolution in the characteristic mass (\mstar\
$\approx 10^{10.65}$\msun) or the slope at low masses ($\alpha \approx
-1.5$) at $0.2<z<2$ in our best-fit double-\sch\ parameterizations.
We also observe evidence of
multiple-component behavior in the star-forming and \qui\ SMFs
independently \citep[see also ][]{Drory09,Gilbank11}.
It is important to note that the low-mass
end of the SMF is dominated by star-forming galaxies with very blue
colors at all \z s. Such galaxies may be subject to systematic
uncertainties in their redshift and mass estimates; while our
photometric \z s appear to be well-constrained (see Fig. \ref{zdist}
and Appendix), spectroscopic confirmation is necessary.

We also examine the growth in the SMFs of the star-forming and
\qui\ populations.  We find that the SMF of star-forming galaxies
increases moderately with cosmic time, by 1.5 - 2.5\x\ since $z \sim
2$, but that the shape of the SMF does not change strongly. These
results are consistent with previous work which has generally found
that the star-forming SMF evolves relatively weakly with redshift
\citep{Bell07,Pozzetti10,Brammer11,Muzzin13}.  For \qui\ galaxies we
observe much more rapid growth in number density, and also a change in
shape of the SMF.  From $z = 2$ to $z = 0$ we find a \twid
6\x\ increase at masses $> 10^{10}$\msun\ and \twid 15 -
30\x\ increase at masses $< 10^{10}$\msun.

Finally, we calculate the evolution of the cosmic stellar mass density
at $z < 3$ integrated between $9 <$ \logm\ $< 13$. 
We compare our results to measurements from UltraVISTA
\citep{Ilbert13,Muzzin13}, which covers a much larger area but at a
much shallower depth, as well as measurements from \citet{Santini12},
which reach a similar depth but over \twid 1/10$^{\rm th}$ of our
survey area. 
Overall, we find good agreement with \citet{Santini12} and
\citet{Ilbert13} at all \z s. Results at $1.5 < z < 2.5$ from
\citet{Muzzin13}, however, are less than what we find.
We also compare to the inferred mass density from \citet{Reddy09}
derived from a rest-frame UV-selected galaxy sample 
corrected for incompleteness.  From this corrected SMF \citet{Reddy09}
measure a value for the cosmic stellar mass density at $1.9 < z < 3.4$
that is similar to ours, despite the use of very different types of data
and different methods.


\acknowledgements

We would like to thank the Mitchell family for their 
continuing support and in particular the late George P. Mitchell whose 
vision and commitment to science and astronomy leaves a lasting 
legacy.
We would also like to thank the Carnegie Observatories and the Las 
Campanas Observatory for providing the facilities and support 
necessary to make the ZFOURGE survey possible. 
This work was supported by the National Science Foundation grant
AST-1009707.
R.F.Q. acknowledges support provided by NASA through Hubble Fellowship
grant \#51279.01 awarded by the Space Telescope Science Institute,
which is operated by the Association of Universities for Research in
Astronomy, Inc., for NASA, under contract NAS 5-26555.

\appendix
\label{appendix}

The steepening of the slope in the SMF at \logm\ $< 10$ is caused by
an excess of low-mass galaxies relative to the expectations from
single-\sch\ functions fit at higher masses.
However, a large number of false detections can produce an artificial
excess. Thus, we have taken careful measures to remove false objects from
our final sample. We remove all detections on or near diffraction
spikes or other stellar aberrations via visual inspection. These types
of interlopers represent roughly 1\% of our final sample.
Spurious detections that arise from noise spikes and extended halos of
nearby galaxies require a more sophisticated approach. For each object
we measure fluxes in two apertures of radii 0.8" and 0.2'' in the
detection image ($H_{160}$). Since true objects are centrally
concentrated the ratio of these fluxes acts as a good
discriminant. After testing, we found that a value of $f_{0.8"} /
f_{0.2"} > 7$ is a reasonable threshold for flagging spurious
detections which account for \twid 10\% of our final sample.
In Figure \ref{seds} we show diagnostic figures for a random subsample of five
star-forming and five \qui\ galaxies at $1.0 < z < 2.5$ that are with 0.5
dex of our calculated mass-completeness limits to help show that our
final sample is not measurably contaminated by false detections.

As can be seen, the low-mass galaxies tend to have very blue colors,
and the photometric redshifts are driven primarily by weak Balmer
breaks and the presence of emission lines in the medium-band
filters. Figure \ref{seds} includes plots of the redshift probability
density calculated using EAZY, which suggest that the redshifts
are quite well-constrained. However we do not rule out that there may
be larger systematic uncertainties for such galaxies, and
spectroscopic confirmation of a significant sample would be
beneficial. 

\begin{figure}[t]
\epsfig{ file=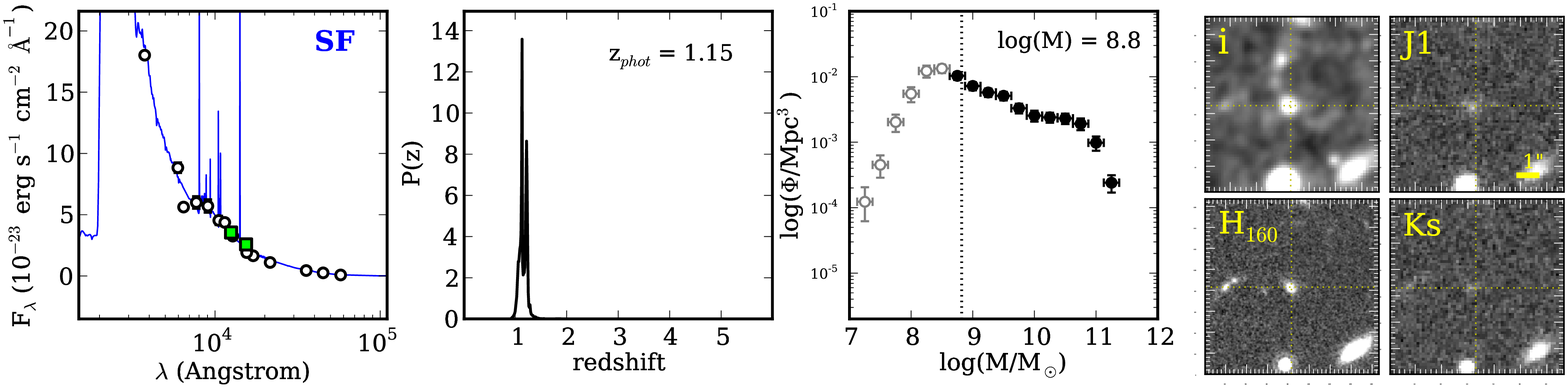, width=0.97\linewidth }
\epsfig{ file=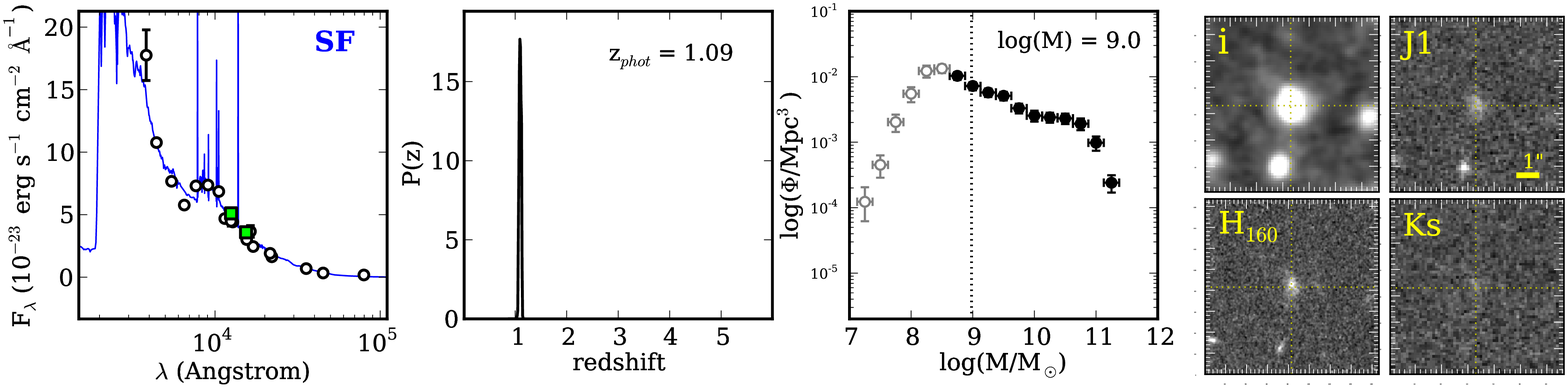, width=0.97\linewidth }
\epsfig{ file=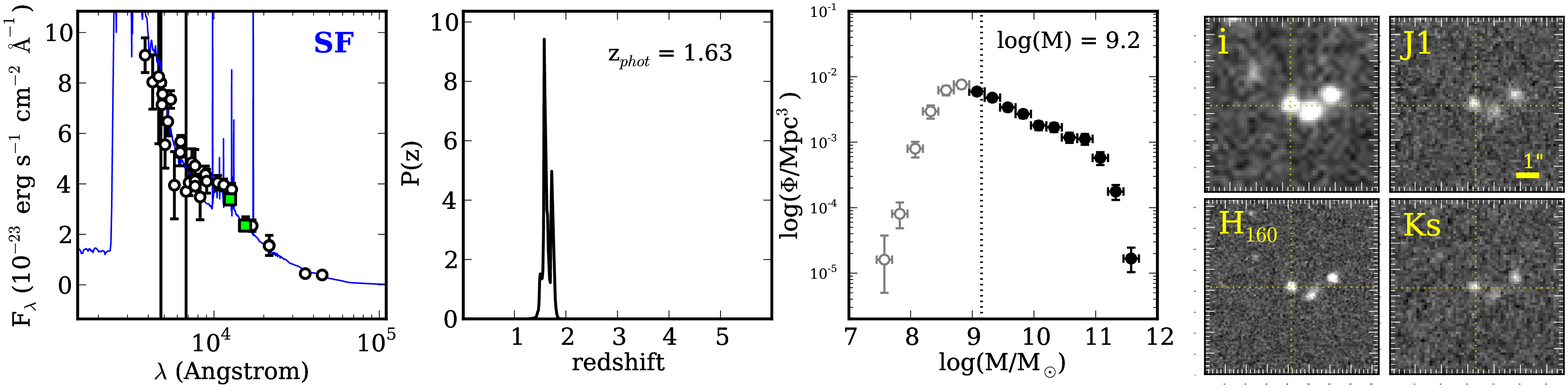, width=0.97\linewidth }
\epsfig{ file=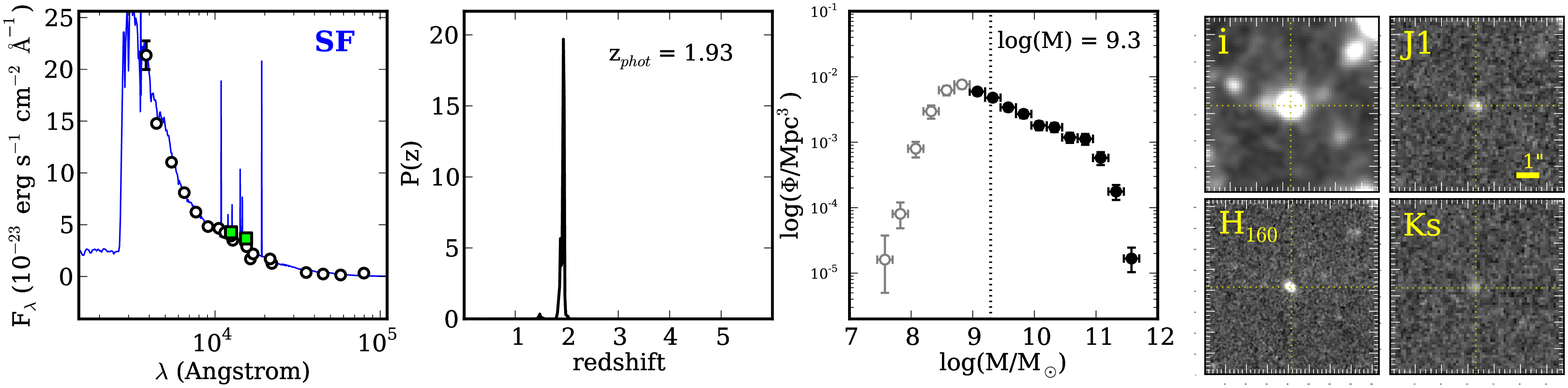, width=0.97\linewidth }
\epsfig{ file=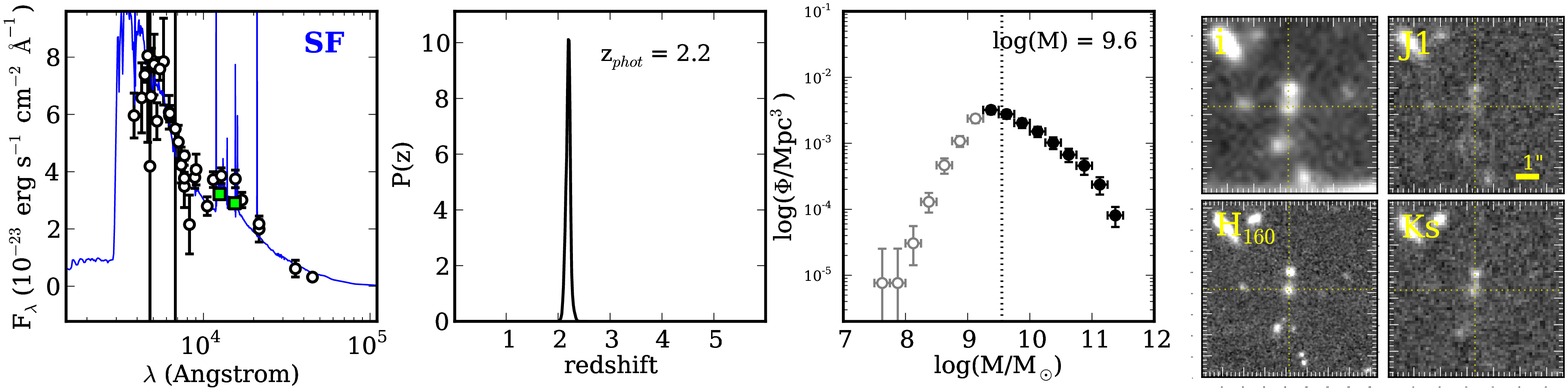, width=0.97\linewidth }
\caption{\figtxt  Diagnostic figures for five star-forming
  and five \qui\ galaxies at $1 < z < 2.5$ that are within 0.5 dex of their
  respective mass-completeness limit. For each galaxy we show (1) the
  measured 0.3 -- 8\um\ photometry with best-fit SED from {\tt EAZY}, green points correspond
  to $J_{125}$ and $H_{160}$ from CANDELS (2) the corresponding \z\
  probability density from {\tt EAZY}, (3) the SMF that the
  galaxy contributes to with a dotted line to indicate the mass of the
  galaxy itself and (4) thumbnails in the $I$-band, $J1$-band, $H_{160}$-band
  and $K_s$-band. Gray open symbols in the SMF panels show
  measurements below our adopted completeness limit at the given \z\
  range.
}
\label{seds}
\end{figure}
\setcounter{figure}{12}
\begin{figure}[t]
\epsfig{ file=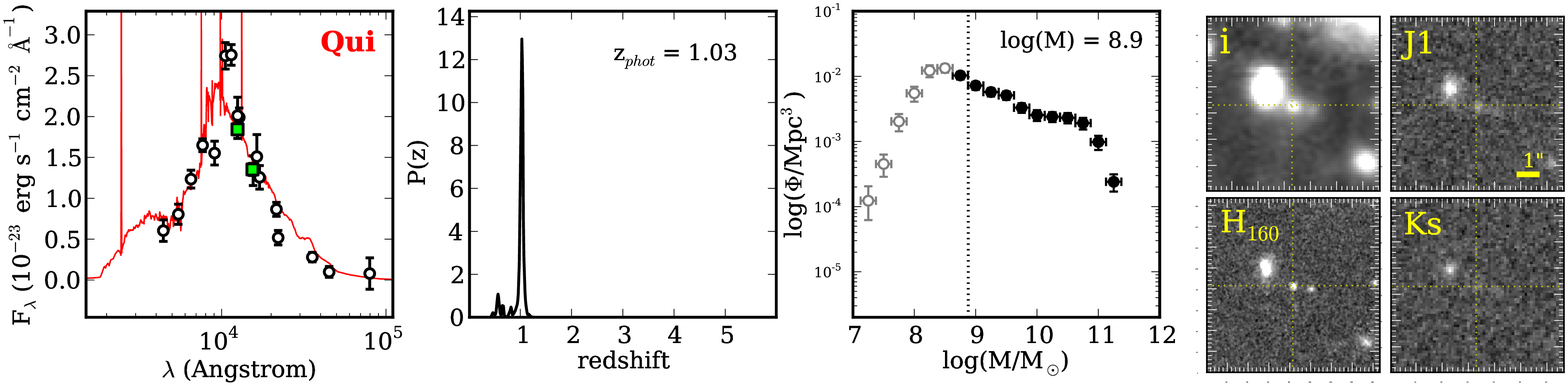, width=0.97\linewidth }
\epsfig{ file=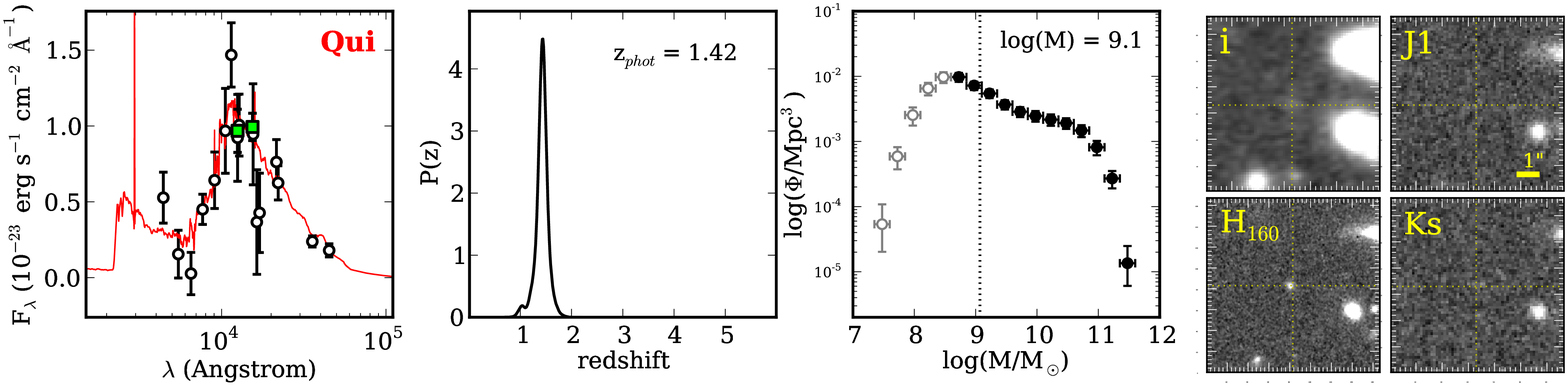, width=0.97\linewidth }
\epsfig{ file=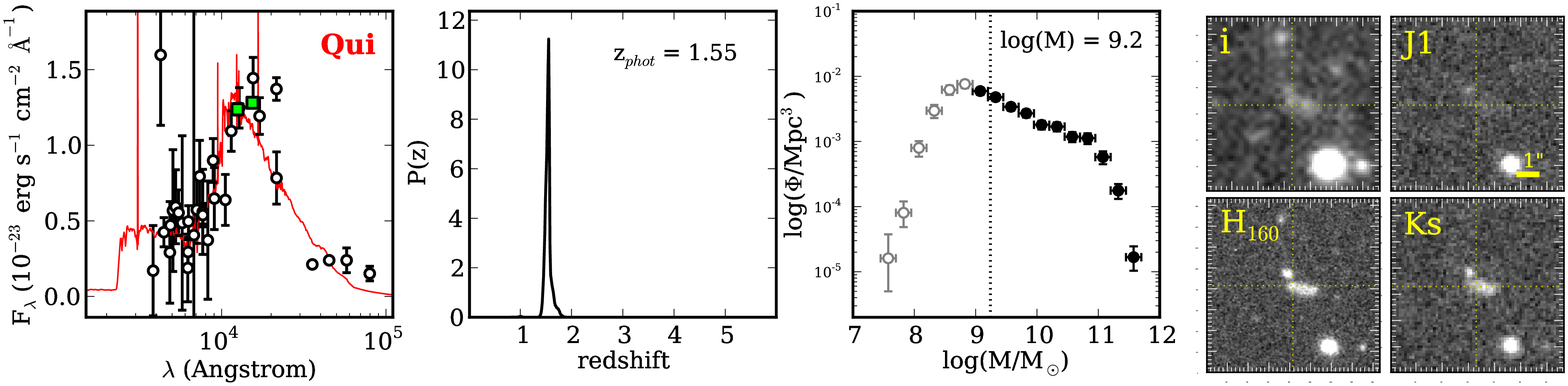, width=0.97\linewidth }
\epsfig{ file=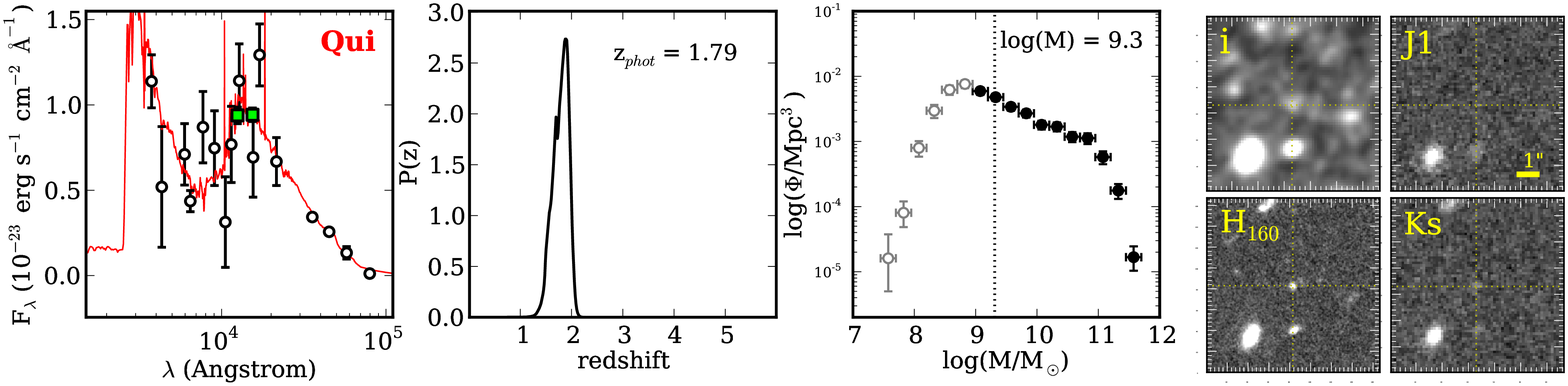, width=0.97\linewidth }
\epsfig{ file=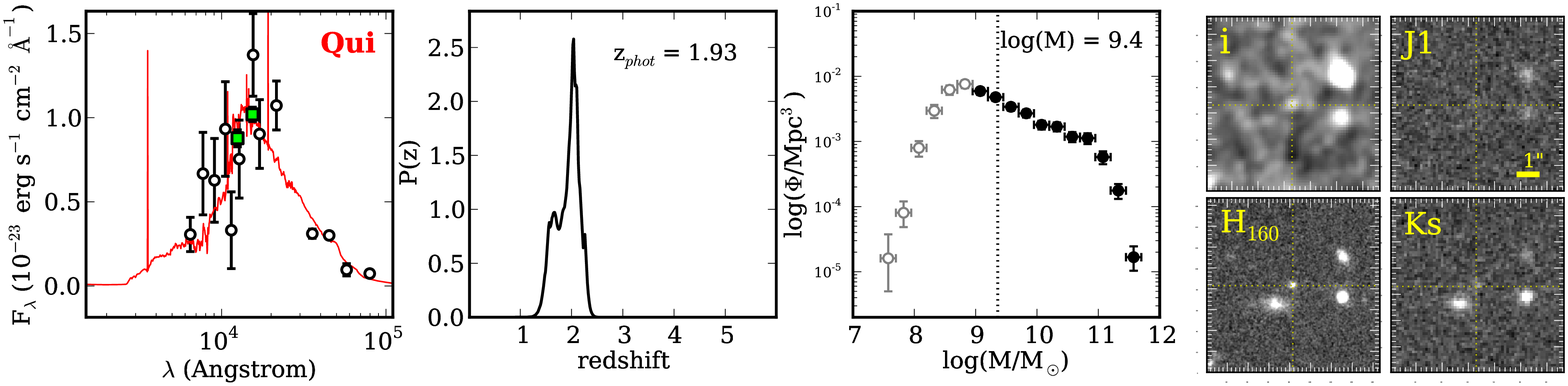, width=0.97\linewidth }
\caption{\figtxt Continued }
\end{figure}

\nocite{*}
\bibliographystyle{apj}
\bibliography{bibliography}

\end{document}